\documentclass[aps,showpacs]{revtex4}

\usepackage{amsmath,amsfonts}
\usepackage{graphicx}

\newcommand{\sgn}{\mathop{\rm{sgn}}}
\newcommand{\rmd}{\mathop{\rm{d}}}

\begin{document}

\title{Phantom cosmology as a scattering process} 

\author{Marek Szyd{\l}owski}
\email{uoszydlo@cyf-kr.edu.pl}
\affiliation{Astronomical Observatory, Jagiellonian University,
Orla 171, 30-244 Krak{\'o}w, Poland}
\affiliation{Mark Kac Complex Systems Research Centre, Jagiellonian University, 
Reymonta 4, 30-059 Krak{\'o}w, Poland}
\author{Orest Hrycyna}
\email{hrycyna@kul.lublin.pl}
\affiliation{Department of Theoretical Physics, Faculty of Philosophy, 
Catholic University of Lublin, Al. Rac{\l}awickie 14, 20-950 Lublin, Poland}
\author{Adam Krawiec}
\email{uukrawie@cyf-kr.edu.pl}
\affiliation{Institute of Public Affairs, Jagiellonian University, 
Rynek G{\l}{\'o}wny 8, 31-042 Krak{\'o}w, Poland}
\affiliation{Mark Kac Complex Systems Research Centre, Jagiellonian University, 
Reymonta 4, 30-059 Krak{\'o}w, Poland}


\begin{abstract}
We study the general chaotic features of dynamics of the phantom field 
modelled in terms of a single scalar field conformally coupled to gravity. 
We demonstrate that the dynamics of the FRW model with dark energy in the form 
of phantom field can be regarded as a scattering process of two types: 
multiple chaotic and classical non-chaotic. It depends whether the 
spontaneously symmetry breaking takes place. In the first class of models with 
the spontaneously symmetry breaking the dynamics is similar to the Yang-Mills 
theory. We find the evidence of a fractal structure in the phase space of 
initial conditions. We observe similarities to the phenomenon of a multiple 
scattering process around the origin. In turn the class of models without the 
spontaneously symmetry breaking can be described as the classical non-chaotic 
scattering process and the methods of symbolic dynamic are also used in this 
case. We show that the phantom cosmology can be treated as a simple model with 
scattering of trajectories which character depends crucially on a sign of a 
square of mass. We demonstrate that there is a possibility of chaotic behavior 
in the flat Universe with a conformally coupled phantom field in the system 
considered on non-zero energy level. We obtain that the acceleration is a 
generic feature in the considered model without the spontaneously symmetry 
breaking. We observe that the effective EOS coefficient oscillates and then 
approach to $w=-1$. 
\end{abstract}


\maketitle

\section{Introduction}

The observations of distant supernovae \cite{Perlmutter:1998np,Riess:1998cb} 
give the evidence that our Universe is undergoing accelerated expansion in the 
present epoch. In principle, there are two different approaches to treat this 
phenomenon. In the first approach it is postulated that there is some unknown 
exotic matter which violates the strong energy condition $\rho + 3p \geq 0$, 
where $p$ is the pressure and $\rho$ is the energy density of perfect fluid. 
This form of matter is called dark energy. In the past few years different 
scalar field models like quintessence and more recently the tachyonic scalar 
field have been conjectured for modelling the dark energy in terms of 
sub-negative pressure $p > - \rho$. A scalar field with super-negative pressure 
$p < - \rho$ called a phantom field can formally be obtained by switching the 
sign of the kinetic energy in the Lagrangian for a standard scalar field. For 
example in the Friedmann-Robertson-Walker (FRW) model the phantom field 
minimally coupled to a gravity field leads to $\rho + p = - {\dot{\psi}}^{2}$, 
where $\rho_{\psi} = -\frac{1}{2} \dot{\psi}^{2} + V(\psi)$, $p_{\psi} = 
-\frac{1}{2} \dot{\psi}^{2} - V(\psi)$, and $V(\psi)$ is the phantom potential. 
Such a field was called the phantom field by Caldwell \cite{Caldwell:1999ew} 
who proposed it as a possible explanation of the observed acceleration of the 
current Universe when $\Omega_{\rm{m},0} \gtrsim 0.2$. Note that a coupling 
to gravity in the quintessence models was also explored \cite{Uzan:1999ch}. 

The second approach called the Cardassian expansion scenario has recently been 
proposed by Freese and Lewis \cite{Freese:2002sq} as an alternative to dark 
energy in order to explain the current accelerated expansion of the Universe. 
In this scenario the Universe is flat and matter dominated but the standard FRW 
dynamics is modified by the presence of an additional term $\rho^{n}$ such that 
$3H^{2} = \rho_{\rm{eff}} = \rho + 3B \rho^{n}$, where $H = (d\ln{a})/dt$ 
is the Hubble parameter; and $a$ is the scale factor. However, let us note 
that this additional term can be interpreted as a phantom field modelled by 
the equation of state $p = p(\rho) = [n(1+ \gamma) - 1] \rho$, where 
$\rho = \rho_{\rm{m},0} a^{-3(1+ \gamma)}$. Therefore for dust matter we 
obtain $p = (n-1) \rho$, and $n<0$ leads to the phantom field. The Cardassian 
expansion with $n<0$ which can be interpreted as the phantom fluid effect. 

Phantom fields lead to the super-accelerated expansion of the Universe, i.e. 
$\rmd H/\rmd t > 0$. The simplest models describe this field in terms of 
minimally coupled real scalar field with the negative kinetic energy term
$-\frac{1}{2} \dot{\psi}^2$ \cite{Caldwell:1999ew,Caldwell:2003vq,Carroll:2003st}. 
It is interesting that phantom fields are also present in string theories 
\cite{Mersini-Houghton:2001su,Bastero-Gil:2001nu,Frampton:2002tu}.
and arise as a phenomenological description of quantum effects of particle 
production in terms of bulk viscosity \cite{Barrow:1988yc}. Because the 
phantom fields violate the Lorentz invariance condition most physicists 
believe that the phantoms open the doors on new physics \cite{Visser:2003yf}.
The investigation of the theoretical possibility to describe dark 
energy in terms of phantom field was the subject of many papers 
\cite{Parker:1999td,Boisseau:2000pr,Schulz:2001yx,Faraoni:2001tq,Onemli:2002hr,
Hannestad:2002ur,Melchiorri:2002ux,Hao:2003ww,Lima:2003dd,Singh:2003vx,
Dabrowski:2003jm,Hao:2003th,Johri:2003rh,Piao:2003ty,Alam:2003fg,Nojiri:2003vn,
Elizalde:2004mq,Sami:2003xv,Gannouji:2006jm}. The review on dark energy models 
were presented by Copeland et al.\ \cite{Copeland:2006wr}. 

In the paper by Dabrowski et al.\ \cite{Dabrowski:2006dd} it is considered the 
quantization of phantom field via the Wheeler-DeWitt equation in quantum 
cosmology. They showed that quantum effects give rise to avoiding the big-rip 
singularity. Also some other basic problems in cosmology like the problem of 
direction of time can be solved.  

If we confront the phantom field model with the observation of SNIa data we 
obtain that it is the best candidate together with the $\Lambda$CDM model for 
the substantial form of dark energy \cite{Choudhury:2003tj,Godlowski:2005tw,
Szydlowski:2006pz,Szydlowski:2006ay}.

It is interesting to investigate new dynamical effects created in the presence 
of phantom fields. The main goal of this paper is to model the phantom fields 
in terms of scalar fields with a potential rather than in terms of the 
barotropic equation of state. For the latter case the dynamics is regular and 
can be represented on the two-dimensional phase space \cite{Szydlowski:2004jv}. 

Usually phantom fields are modelled in terms of minimally coupled scalar fields 
The case of minimally coupled to gravity phantom fields was analyzed in the 
context of existence of periodic solutions \cite{Giacomini:2006ak}. They 
demonstrated that the dynamics is trivial in a sense of nonexistence of periodic 
solutions. Faraoni used the framework of phase space for investigating the 
dynamics of phantom cosmology, late-time attractors, and their existence for 
different shapes of potential \cite{Faraoni:2005gg}. It was showed that 
dynamics of the flat FRW universe can be reduced to the form of a two-dimensional 
dynamical system on a double sheeted phase space. It is a simple consequence 
of an algebraic equation for $\dot{\psi}$ which can be expressed as a function of
the Hubble parameter $H$ and $\psi$. In this case there is no place for chaotic 
behavior of trajectories in the phase space $(H,\psi,\dot{\psi})$ 
\cite{Faraoni:2006sr}. The exit on the inflationary epoch and bounce was 
showed in the flat FRW universe with two interacting phantom scalar fields 
\cite{Dzhunushaliev:2006xh}. In the closed FRW cosmological model with minimally 
coupled scalar field there appears transient chaos which has character 
of the scattering process \cite{Toporensky:2005uk}. The scattering process takes 
place during the bounce --- a transition from a contracting to expanding universe.
In this context the language of symbolic dynamics and topological entropy was 
used \cite{Cornish:1997ah,Page:1984qt,Kamenshchik:1998ix}.

The conformally coupled scalar field are also of interest. The significance 
of long-wavelength modes in the WKB approximation of a conformally coupled 
scalar field was analyzed in the inflationary scenario \cite{Jankiewicz:2005tm}. 

While in this paper we concentrate on the conformal coupled scalar fields, it 
is worth to mention the works about the case of the generic (non-minimal and 
non-conformal) coupling between a phantom field and gravitation. Faraoni 
\cite{Faraoni:2000gx} discusses different motivations for choosing $\xi \ne 0$ 
(both positive and negative) from the theoretical point of view and from 
observational constraints. The inflation and quintessence with the non-minimal 
coupling were studied in the context of the formulation of some necessary 
conditions for acceleration of the universe \cite{Faraoni:2000wk,Faraoni:2000nt} 
(see also \cite{Bellini:2002zr}). Faraoni also pointed out that the 
non-minimally coupling term different from the conformal coupling value 
($\xi = 1/6$) spoils the equivalence principle of general relativity. 

It is interesting that some observational constraints on the non-minimal
coupling (possitive or negative) can be found from CMB observations. 
Tsujikawa and Gumjjudpai \cite{Tsujikawa:2004my} studied this bounds for a 
special power-law potential. The dynamics with the exponential form of the 
potential of a scalar field was also studied \cite{Tsujikawa:2000tm}. 
The two-field inflation models with negative non-minimal coupling and hybrid 
inflation are also interesting in the context of the large scale curvature 
perturbation \cite{Tsujikawa:2002nf}. In turn the constraints on the 
ratio of the self-coupling and non-minimal coupling constant can be obtained 
during the inflation period \cite{Noh:2000kr}.

In this paper we ask what kind of dynamics can be expected from the FRW model 
with phantom field. It is well known that the standard FRW model reveals some 
complex dynamics. The detailed studies gave us a deeper understanding of 
dynamical complexity and chaos in cosmological models and resulted in 
conclusion that complex behavior depends on the choice of a time 
parameterization or a lapse function in general relativity 
\cite{Misner:1969hg,Belinsky:1970ew,Cornish:1995mf}. Castagnino et al. 
\cite{Castagnino:1999wd} showed that dynamics of the closed FRW models with a 
conformally coupled massive scalar field is not chaotic if considered in the 
cosmological time. The same model was analyzed in the conformal time by 
Calzetta and Hasi \cite{Calzetta:1992bv} who presented the existence of chaotic 
behavior of trajectories in the phase space. Motter and Letelier 
\cite{Motter:2002jj} explained that this contradiction in the results is 
obtained because the system under consideration is non-integrable. Therefore we 
can speak about complex dynamics in terms of nonintegrability rather than 
deterministic chaos. The significant feature is that nonintegrability is an 
invariant evidence of dynamical complexity in general relativity and cosmology 
\cite{Maciejewski:2000}. 

We can find many analogies between models with spontaneously symmetry breaking 
and the Yang-Mills systems. Problems of chaotic behavior in the Yang-Mills 
cosmological models have been investigated by Barrow and his collaborators. 
Chaos from Yang-Mills can only occur in an anisotropic universe but it occurs 
for some arbitrarily small anisotropy (see the study of Bianchi I 
\cite{Barrow:1997sb} and other Bianchi types \cite{Jin:2004vh}). However, when 
a perfect fluid is added things change in an interesting way that contrast with 
the situation with a magnetic field \cite{Barrow:2005df}.

For the FRW model with phantom it can be shown that there is a monotonous 
function along its trajectories and it is not possible to obtain the Lyapunov 
exponents or construct the Poincar{\'e} sections. Therefore it is useful to 
study the nonintegrability of the phantom system and set it in a much stronger 
form by proving that the system does not possess any additional and independent 
of Hamiltonian first integrals, which are in the form of analytic or meromorphic 
functions. Of course, it is not the evidence of sensitive dependence of 
solution on a small change of initial conditions. However, it is the possible 
evidence of complexity of dynamical behavior formulated in an invariant way 
\cite{Szydlowski:2004jv}. 

The notion of deterministic chaos is a controversial issue in the Mixmaster 
models. The value of the numerically computed maximal Lyapunov exponent for 
those systems depends on the time parameterization used \cite{Rugh:1990aq}. 
However, the existence of fractal structures in the phase space provides 
the coordinate independent signal of chaos in cosmology as it was shown by 
Cornish and Levin \cite{Cornish:1995mf}. In particular, they found that the 
Bianchi IX has a form of chaotic scattering. The short scattering periods 
intermittent integrable motion and evolution of the system is chaotic. It is 
similar to a pin-ball machine.

The main goal of this paper is to show that dynamics of phantom cosmology can 
be treated as a scattering process. Only if the spontaneously symmetry breaking 
is admitted this process has chaotic character. Evidences of dynamical behavior 
are studied in tools of symbolic dynamics, fractal dimension, and analytically
the Toda-Brumer-Duff test.

\section{Hamiltonian dynamics of phantom cosmology}

We assume the model with FRW geometry, i.e., the line element has the form
\begin{equation}
\rmd s^{2} = -\rmd t^{2} + a^{2}(t)[\rmd\chi^{2} + f^{2}(\chi)(\rmd\theta^{2}
+ \sin^{2}{\theta}\rmd\varphi^{2})],
\label{eq:1}
\end{equation}
where
\begin{equation}
f(\chi) = \left \{
\begin{array}{lll} 
\sin{\chi},  & 0 \leq \chi \leq \pi     & k=+1 \\
\chi,        & 0 \leq \chi \leq \infty  & k=0  \\
\sinh{\chi}, & 0 \leq \chi \leq \infty  & k=-1 
\end{array} 
\right.
\label{eq:2}
\end{equation}
$k=0,\pm 1$ is the curvature index, $0 \leq \varphi \leq 2\pi$ and 
$0 \leq \theta \leq \pi$ are comoving coordinates, $t$ stands for the 
cosmological time.

It is also assumed that a source of gravity is the phantom scalar field 
$\psi$ with a generic coupling to gravity. The gravitational dynamics is 
described by the standard Einstein-Hilbert action
\begin{equation}
S_{g}=\frac{1}{2}m_{p}^{2}\int \rmd^{4}x \sqrt{-g}(R-2\Lambda),
\label{eq:3}
\end{equation}
where $m_{p}^{2}=(8\pi G)^{-1}$; for simplicity and without lost of generality 
we assume $4\pi G/3=1$. The action for the matter source is
\begin{equation}
S_{\rm{ph}} = -\frac{1}{2}\int \rmd^{4}x
\sqrt{-g}(-g^{\mu\nu}\psi_{\mu}\psi_{\nu} - \xi R\psi^{2} + 2 U(\psi)).
\label{eq:4}
\end{equation}
Let us note that the formal sign of $||\psi||^{2}$ is opposite to that which 
describes the standard scalar field as a source of gravity, where $U(\psi)$ 
is a scalar field potential. We assume
\begin{equation}
U(\psi) = \frac{1}{2}m^{2}\psi^{2} + \frac{1}{4}\lambda\psi^{4}
\label{eq:5}
\end{equation}
and that conformal volume $\int d^{3}x$ over the spatial 3-hypersurface is 
unity. $\xi$ is the coupling constant of the scalar field to the Ricci scalar
\begin{equation}
R=6\left(\frac{\ddot{a}}{a}+ \frac{\dot{a}^2}{a^2} + \frac{k}{a^{2}}\right)
\label{eq:6}
\end{equation}
where a dot means the differentiation with respect to the cosmic time $t$.

If we have the minimally coupled scalar field then $\xi=0$. We assume a 
non-minimal coupling of the scalar field $\xi \neq 0$. 

The dynamical equation for phantom cosmology in which the phantom field is 
modelled by the scalar field with an opposite sign of the kinetic term in 
action can be obtained from the variational principle 
$\delta(S_{g}+S_{\rm{ph}})=0$. After dropping the full derivatives with 
respect to time we obtain the dynamical equation for phantom cosmology from 
variation $\delta(S_{g}+S_{\rm{ph}})/\delta g=0$ as well as the dynamical 
equation for field from variation $\delta(S_{g}+S_{\rm{ph}})/\delta \psi=0$ 
\begin{equation}
\ddot{\psi}+3H\dot{\psi} - \frac{\rmd U}{\rmd\psi}+\xi R \psi = 0.
\label{eq:7}
\end{equation}
It can be shown that for any value of $\xi$ the phantom behaves like some 
perfect fluid with the effective energy $\rho_{\psi}$ and the pressure 
$p_{\psi}$ in the form which determines the equation of state factor
\begin{equation}
w_{\psi}=\frac{-\frac{1}{2}\dot{\psi}^{2}-U(\psi)
+ \xi[2H(\psi^{2})\dot{\ } + (\psi^{2})\ddot{\ }] 
+ \xi \psi^{2}(2\dot{H}+3H^{2})}{-\frac{1}{2}\dot{\psi}^{2}+U(\psi)
- 3\xi H[H\psi^{2}+(\psi^{2})\dot{\ }]} \equiv \frac{p_{\psi}}{\rho_{\psi}}.
\label{eq:8}
\end{equation}
Formula~(\ref{eq:8}) differs from its counterpart for the standard scalar 
field \cite{Gunzig:2000ce} by the presence of a negative sign in front of 
the term $\dot{\psi}^{2}$.

In equation~(\ref{eq:8}) the second derivative $(\psi^{2})\ddot{\ }$ in the 
expression for the pressure can be eliminated and then we obtain
\begin{equation}
p_{\psi}=\left(-\frac{1}{2}+2\xi\right)\dot{\psi}^{2}
-\xi H\big(\psi^{2}\big)\dot{\ }-2\xi\big(6\xi-1\big)\dot{H}\psi^{2}
-3\xi\big(8\xi-1\big)H^{2}\psi^{2}-U(\psi)+2\xi \psi \frac{\rmd U}{\rmd\psi}.
\label{eq:9}
\end{equation}
Of course such perfect fluid which mimics the phantom field satisfies the 
conservation equation
\begin{equation}
\dot{\rho}_{\psi}+3H(\rho_{\psi}+p_{\psi})=0.
\label{eq:10}
\end{equation}
We can see that complexity of a dynamical equation should manifest by 
complexity of $w_{\psi}$.

Let us consider the FRW quintessential dynamics with some effective energy 
density $\rho_{\psi}$ given in equation~(\ref{eq:8}). This dynamics can be 
reduced to the form like of a particle in a one-dimensional potential 
\cite{Szydlowski:2003cf} and the Hamiltonian of the system is 
\begin{equation}
\mathcal{H}(a',a)=\frac{(a')^{2}}{2}+V(a) \equiv 0,\qquad 
V(a)=-\rho_{\psi}a^{4}
\label{eq:11}
\end{equation}
where a prime means the differentiation with respect to the conformal time 
$\eta$. 

The trajectories of the system lie on the zero energy level for flat and 
vacuum models. Note that if we additionally postulate the presence of 
radiation matter for which $\rho_{r} \propto a^{-4}$ then it is equivalent 
to consider the Hamiltonian on the level $\mathcal{H}=E=\rm{const}$. 
Of course the division on kinetic and potential parts has only a conventional 
character and we can always translate the term containing $\dot{\psi}^{2}$ 
into a kinetic term.

The dynamics of the model is governed by the equation of motion (\ref{eq:7}), 
which is equivalent to the conservation condition (\ref{eq:10}) and the 
acceleration condition
\begin{equation}
\label{eq:12}
\frac{\ddot{a}}{a} = - \rho_{\psi}(1 + 3w_{\psi}).
\end{equation}
This equation admits the generalized Friedmann first integral which assumes 
the following form
\begin{equation}
- \frac{1}{2} \dot{\psi}^{2} + U(\psi) - 3 \xi H^{2} \psi^{2} - 3 \xi H
(\psi^{2})\dot{\ }  = \frac{1}{2}H^2 - \frac{1}{6}\Lambda - \rho_{r}.
\label{fint}
\end{equation}
If we postulate existence of radiation in model then left hand of this equation can be negative.

Let us consider now both cases of minimally and conformally coupled phantom 
fields. They can model the quintessence matter field in terms of matter 
satisfying the equation of state $p_{\psi} = w_{\psi} \rho_{\psi}$.

\subsection{Minimally coupled phantom fields}

For minimally coupled phantom fields ($\xi=0$) the function of energy takes 
the form
\begin{equation}
\mathcal{E} = \frac{(a')^{2}}{2} 
+ \frac{1}{2}a^{-2}\left(\phi' a - \phi a' \right)^{2} 
- \frac{1}{2}m^{2}\phi^{2}a^{2} - \frac{\lambda}{4}\phi^{4} 
- \frac{\Lambda}{6}a^{4},
\label{eq:13}
\end{equation}
where $\rho_{\rm{eff}}=-\frac{1}{2}\dot{\psi}^{2}+U(\psi)$, 
$V=-\rho_{\rm{eff}} a^{4}$, $\mathcal{H}=\frac{1}{2}\dot{a}^{2}
+V(a,\psi,\dot{\psi})$, $\phi=a\psi$,
$U(\psi)=\frac{1}{2}m^{2}\psi^{2}+\frac{1}{4}\lambda\psi^{4}$ is assumed.

\subsection{Conformally coupled phantom fields}

For conformally coupled phantom fields we put $\xi=1/6$ and rescale the field 
$\psi \rightarrow \phi = \psi a$. Then the energy function takes the following 
form for a simple mechanical system with a natural Lagrangian function 
$\mathcal{L}=\frac{1}{2}g_{\alpha\beta}(q^{\alpha})' (q^{\beta})' - V(q)$
\begin{equation}
\mathcal{E} = \frac{1}{2}\left((a')^{2}+(\phi')^{2}\right) 
-  \frac{1}{2}m^{2}\phi^{2}a^{2} - \frac{\lambda}{4}\phi^{4} -
\frac{\Lambda}{6}a^{4}.
\label{eq:14}
\end{equation}
In contrast to the FRW model with conformally coupled scalar field the kinetic 
energy form is positive definite like for classical mechanical systems. The 
general Hamiltonian which represents the special case of a two coupled 
non-harmonic oscillators system is 
\begin{equation}
\mathcal{H}=\frac{1}{2}g^{\alpha\beta}p_{\alpha}p_{\beta}+V(q) 
= \frac{1}{2}(p_{x}^{2}+p_{y}^{2})+Ax^{2}+By^{2}+Cx^{4}+Dy^{4}+Ex^{2}y^{2},
\label{eq:15}
\end{equation}
where $A$, $B$, $C$, $D$, and $E$ are constants.

In order to study the integrability of dynamical systems we use Painlev{\'e}'s 
approach. Painlev{\'e}'s analysis gives necessary conditions for the 
integrability of dynamical systems and it is the most popular integrability 
detector. The recapitulation of Painlev{\'e}'s analysis for system~(\ref{eq:15}) 
was done by Lakshmanan and Sahadevan \cite{Lakshmanan:1993}. They found that 
system~(\ref{eq:15}) passes the Painlev{\'e} test and is integrable in the 
following four cases 
\begin{eqnarray*}
\Lambda =& \lambda \qquad	& m^{2} = 3\Lambda	\\
\Lambda =& \lambda \qquad	& m^{2} = \Lambda	\\
\Lambda =& 8\lambda \qquad	& m^{2} = 3\Lambda	\\
\Lambda =& 16\lambda \qquad	& m^{2} = 6\Lambda.
\end{eqnarray*}
This result is in full agreement with conclusions concerning integrability of 
the two coupled quartic non-harmonic oscillator systems. Therefore, phantom 
cosmology can be considered as coupled quartic non-harmonic oscillators. 
Of course, Painlev{\'e} analysis gives necessary conditions for the 
integrability of dynamical systems. However, there exist whole classes of 
integrable systems which do not possess the Painlev{\'e} property. It means that 
the Painlev{\'e} approach gives over-restrictive conditions for integrability. 
It is obvious that for $m^{2}=0$ the FRW phantom cosmology is integrable 
because of the possibility to separate of variables in the potential. Then 
we have two decoupled quartic non-harmonic oscillators.

The non-integrability of the non-flat FRW model with the scalar field with the 
potential $V(\phi) \propto \phi^2$ was investigated in an analytical way by 
Ziglin \cite{Ziglin:2000}. For a deeper analysis of integrability in terms of 
Ziglin and Morales-Ruiz and Ramis see \cite{Szydlowski:2004jv}. 

It would be useful to compare Hamiltonians for the cosmological model with 
phantom fields and the standard cosmological model with scalar fields. 
Let us consider that $\Lambda = \lambda = 0$ for simplicity of presentation. 
Then for both models we have Hamiltonians
\begin{equation}
\mathcal{H}_{\rm{ph}} = \frac{1}{2}(-p_{a}^{2} - p_{\phi}^{2}) 
+ \frac{k}{2}(\phi^2 - a^2) + \frac{1}{2}m^2 a^2 \phi^2
\label{eq:16}
\end{equation}
and
\begin{equation}
\mathcal{H}_{\rm{FRW}} = \frac{1}{2}(-p_{a}^{2} + p_{\phi}^{2}) 
+ \frac{k}{2}(\phi^2 - a^2) + \frac{1}{2}m^2 a^2 \phi^2.
\label{eq:17}
\end{equation}
If we add a radiation component to the energy momentum tensor, which energy 
density scales like $\rho_{\rm{r}}= \rho_{\rm{r},0}a^{-4}$ then both 
systems should be considered on the constant energy level $\mathcal{H}=\mathcal{E}=
\rho_{\rm{r},0}$ or the constant $\rho_{\rm{r},0}$ can be absorbed by 
the new Hamiltonian $\bar{\mathcal{H}} \equiv \mathcal{H} - \rho_{\rm{r},0}$ 
and $\bar{\mathcal{H}}$ is considered on the zero energy level. 

Let us concentrate on the flat models to analyze the similarities to the 
Yang-Mills systems. While the standard cosmological model is described by the 
Hamiltonian 
\begin{equation}
\mathcal{H}_{\rm{FRW}} = \frac{1}{2}(-p_{a}^{2} + p_{\phi}^{2}) 
+ \frac{1}{2}m^2 a^2 \phi^2.
\label{eq:18}
\end{equation}
the phantom cosmological model is
\begin{equation}
\mathcal{H}_{\rm{ph}} = \frac{1}{2}(-p_{a}^{2} - p_{\phi}^{2}) 
+ \frac{1}{2}m^2 a^2 \phi^2
\label{eq:19}
\end{equation}

The crucial difference between system (\ref{eq:18}) and (\ref{eq:19}) lies 
in the definiteness of their kinetic energy forms. It is indefinite and has 
the Lorentzian signature for the standard model, and it is positive definite 
for the phantom model. As a consequence we obtain that the configuration 
space for the standard model is $\mathbb{R}^2$ whereas the condition 
$\rho_{\rm{r},0} + \frac{1}{2} m^2 \phi^2 a^2 >0$ determines the domain 
of the configuration space admissible for motion for the phantom model. 
Note that if $m^2 <0$ (the model with the spontaneously symmetry breaking) then 
this domain is bounded by four hyperbolas in every quarter of plane 
$(a,\phi)$. The same situation can be obtained for the model without 
the spontaneously symmetry breaking ($m^2 > 0$) and dark radiation 
($\rho_{\rm{r},0} < 0$). 

The flat phantom model with the spontaneously symmetry breaking is well known as 
the Yang-Mills systems which have been analyzed since the pioneering paper by 
Savvidy \cite{Savvidy:1982jk}. For this system the Lyapunov exponents were 
found \cite{Kawabe:1991qy} and the Poincar{\'e} sections were obtained 
\cite{Benettin:1976dz}. The spatially flat universes filled with the Yang-Mills 
fields exhibit chaotic oscillations of these fields \cite{Barrow:1997sb,
Jin:2004vh,Barrow:2005df}.

This system was also investigated by using the Gaussian curvature criterion 
\cite{Toda:1974}. Let us apply this criterion to the non-flat phantom model. 
The potential function takes the following form 
\begin{equation}
V(a,\phi) = - k (\phi^2 - a^2) - \frac{1}{2} m^2 \phi^2 a^2
\label{eq:20}
\end{equation}
According to this criterion the periodic and quasi-periodic orbits appear in 
the domains of the configuration space in which the Gaussian curvature of 
a diagram of the potential function is positive. The line of zero curvature 
separates these domains from the instability regions where the curvature 
is negative. If the total energy of the system $E$ increases the system 
will be in a region of negative curvature for some initial conditions and 
the motion is chaotic. 

Let us consider the dynamical system in an autonomous form in the form 
$(x^{i})' = f^i(x^{j})$ then we linearized around the special solution and 
we obtain equation
\[
(\delta x^{i})' = \frac{\partial f^i}{\partial x^k} \delta x^k
\]
where $\frac{\partial f^i}{\partial x^k}$ builds the Jacobian and $\delta x^k$ 
is the deviation vector connecting the points on two nearby trajectories 
corresponding the same value of parameter $t$. In our case $x^1 = a$, 
$x^2 = a'$, $x^3 = \phi$, $x^4 = \phi'$ then the local instability 
of nearby trajectories are determined by eigenvalues of the Jacobian matrix, 
i.e. the effective potential $V(a,\phi)$. The eigenvalues are
\begin{equation}
\label{eq:21}
\mu_{1,2} = \frac{1}{2} \left( -3H \pm \sqrt{9H^2 + 4\gamma}\right)
\end{equation}
\[
\gamma = \gamma_{1,2} = \frac{1}{2} \left(-V_{\phi \phi} - V_{aa} \pm 
\sqrt{(V_{\phi \phi} + V_{aa})^2 - 4 (V_{\phi \phi} V_{aa} - 
V_{\phi a}^2)}\right)
\]
Therefore the necessary condition for local instability is that at least one 
of the eigenvalues is positive \cite{Jin:2004bf}. These positive values 
decide about the local instability of trajectories. Note that the negative 
sign of the Gaussian curvature of potential $V(a,\phi)$ is the sufficient 
condition of local instability \cite{Toda:1974,Brumer:1976}
\begin{equation}
\label{eq:22}
\sgn K = \sgn ( V_{\phi \phi} V_{aa} - V_{\phi a}^{2}) < 0
\end{equation}
We can see that test~(\ref{eq:22}) of negative curvature adopted in our case 
gives to $K<0$. However it is not a sufficient condition for the chaos. It 
should be pointed out that this criterion has a purely local character in 
contrast to the Lyapunov exponent. Moreover the compactness of a region  
admissible for motion is required for chaos existence. 

It is worthwhile to mention that the Toda criterion is only the measure of 
the local instability of nearby trajectories \cite{Nakazato:1995cn}.
Therefore the presented analysis of the order-chaos transition should be 
combined with the Poincar{\'e} sections and other deeper indicators of 
the chaotic behavior.

\section{Numerical investigations of chaos in phantom cosmology}

To make the numerical analysis we need to distinguish the the model without 
the spontaneously symmetry breaking ($m^2 > 0$) and the model with the 
spontaneously symmetry breaking ($m^2 < 0$). In both cases the system is 
considered on the some energy level.

\subsection{$m^2 >0$}

We would like to stress out why we use the nonintegrability criterion instead 
of standard measure like the Lyapunov exponents or Poincar{\'e} sections. The 
main reason is that in analogy to Castagnino et al.'s work 
\cite{Castagnino:1999wd} we can find a monotonous function along a trajectory. 
This excludes the property of the recurrence the trajectories or topological 
transitivity in the standard Wiggins chaos definition. Then 
$F(\phi,\dot{\phi})$ (or $F(a,\dot{a})$) is a monotonous function of the time 
parameter along the trajectories and trajectories escape to infinity for 
arbitrary initial conditions. In this case it is impossible to construct the 
Poincar{\`e} sections. It is obvious for any curvature and positive values of 
$\Lambda$, $\lambda$, and $m^2$. The phase space as well as the configuration 
space is unbounded in this case. In contrast to classical chaotic systems 
there is invariant compact chaotic set for this system. It does not mean the 
system is non-chaotic as it does not follow the Wiggins definition of chaos. 
The essence of this phenomenon has different nature. The system is 
oversensitive with respect to small change of initial conditions like in 
chaotic scattering processes. 

The typical case of the model is a class of models without the spontaneously 
symmetry breaking
\[
\mbox{class A:} \qquad A=B=C=D=0, \quad E = -1 \mbox{ (or } m^2 > 0).
\]
In this case the domain admissible for motion is unbounded. For the classical 
mechanical systems with chaos there are chaotic sets of trajectories on a 
compact invariant submanifold. However it does not mean that the system with 
an unbounded region admissible for motion is nonchaotic because the essence of 
chaotic behavior is of different nature. The system can be oversensitive with 
respect to small changes of initial conditions---the key ingredient of chaos. 
In this respect the system is similar to chaotic scattering processes 
\cite{Dettmann:1994dj}. Therefore it seems to be natural to use the methods of 
investigation of scattering processes. 

The class A model does not possess the property of sensitive dependence on 
initial conditions. Moreover, there are no fractal structures of basin 
boundaries in the phase space. Because the scattering of trajectories in the 
potential well is present, we deal with the nonchaotic scattering process. In 
more details it will be discussed in Sec.~\ref{sec:4}.

The different chaotic behavior was found in the Einstein-Yang-Mills (EYM) 
color field in the flat Bianchi I spacetime \cite{Barrow:1997sb}. It was 
considered the model with radiation and isotropic curvature in contrast to 
the Mixmaster models. It was showed that the EYM systems are also an example 
of chaotic scattering. Because the Lyapunov exponents are coordinate dependent 
they cannot be used for invariant characterization of chaos. For this aim 
the methods of chaotic scattering are more suitable. They are extremely 
useful in this context because of noncompact phase space (also the 
configuration space)---the major obstacle of the standard analysis of chaos. 
Let us note that the case of their model with flat spacetime is analogous to 
our second case with the cosmological constant.

\subsection{$m^2 <0$}

\begin{figure}[t]
\begin{center}
\includegraphics[scale=0.67]{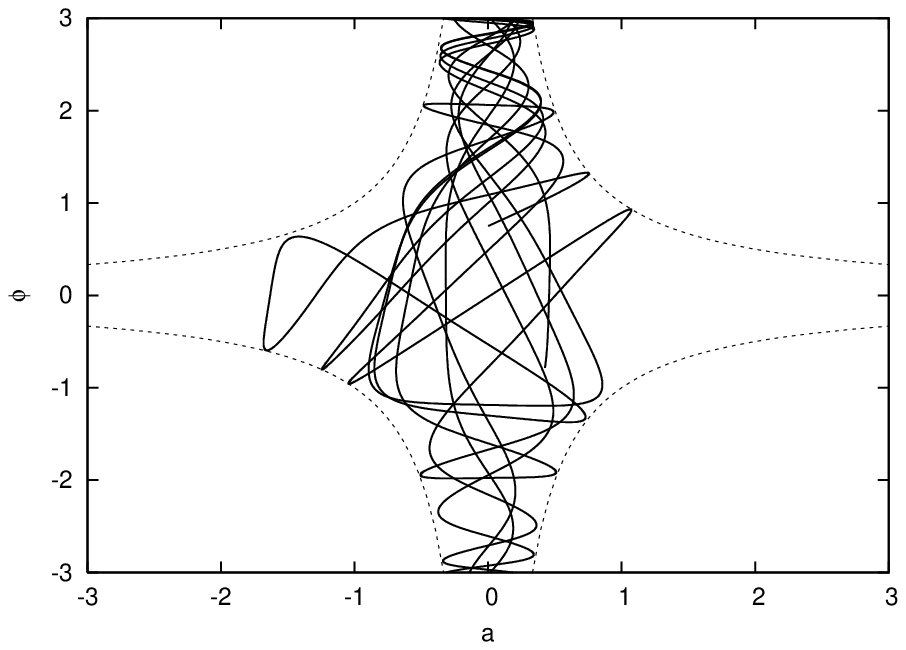}
\includegraphics[scale=0.67]{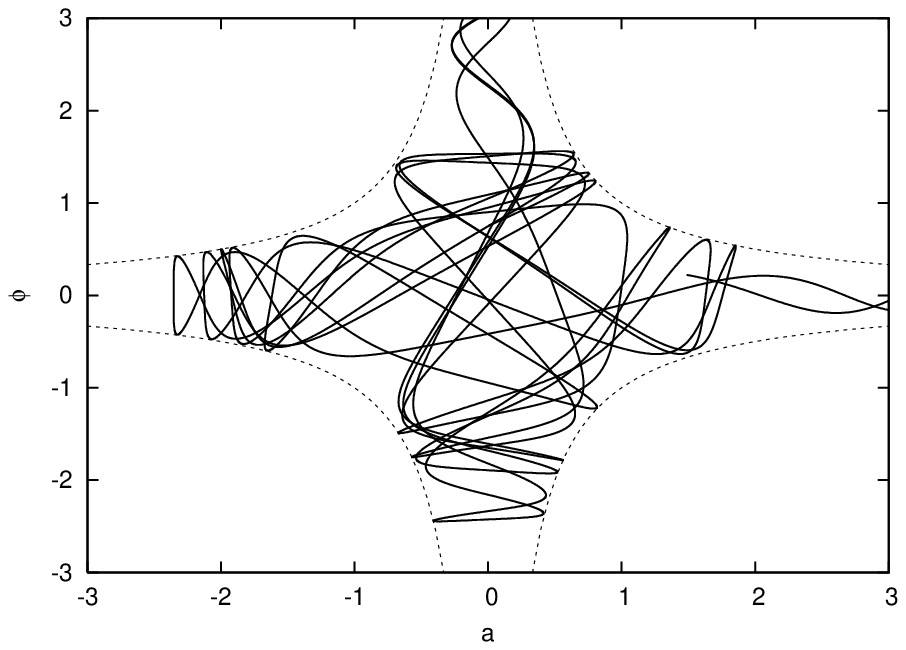}
\end{center}
\caption{Two samples of trajectories for model class I. Initial conditions 
$a_{0}=0$, ${\phi'}_{\! \! 0}=0.779$, $\phi_{0}=0.75$ (left) 
and $\phi_{0}=0.751$ (right), ${a'}_{\! \! 0}>0$ calculated from the 
Hamiltonian constraint.}
\label{fig:1}
\end{figure}

\begin{figure}[t]
\begin{center}
\includegraphics[scale=0.67]{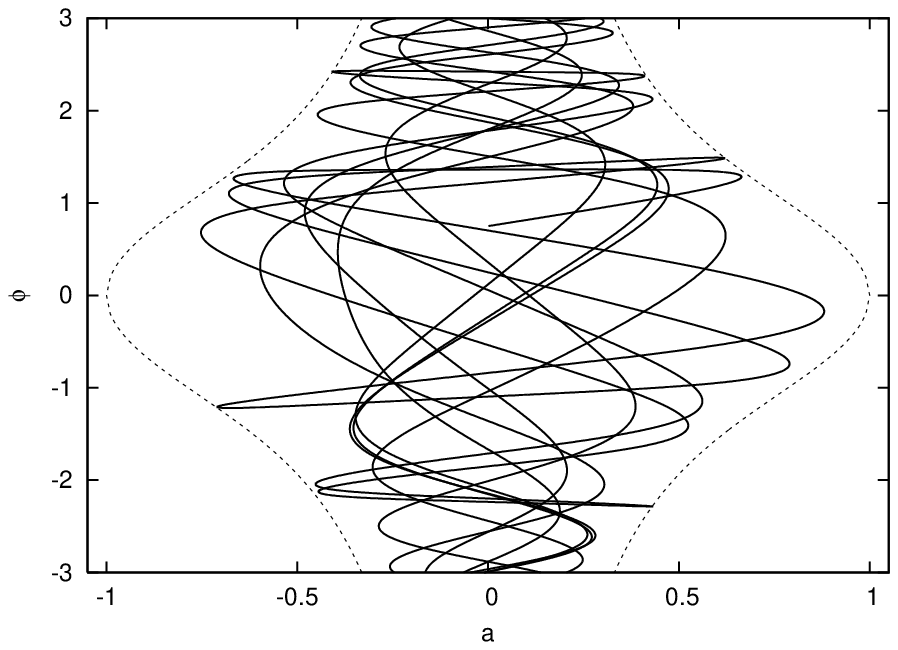}
\includegraphics[scale=0.67]{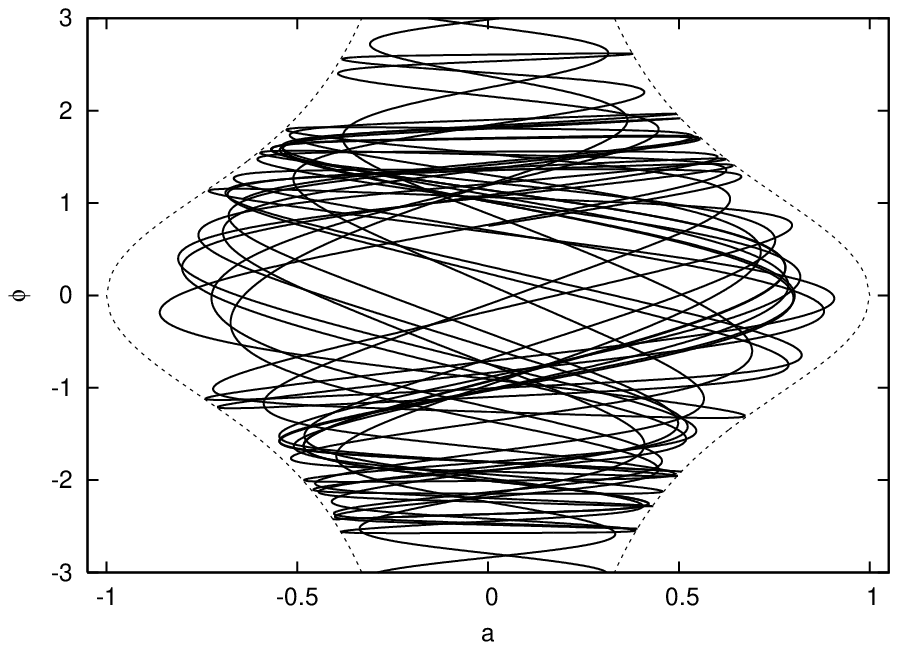}
\end{center}
\caption{Two samples of trajectories for model class II. Initial conditions 
$a_{0}=0$, ${\phi'}_{\! \! 0}=0.779$, $\phi_{0}=0.75$ (left) 
and $\phi_{0}=0.751$ (right), ${a'}_{\! \! 0}>0$ calculated from the 
Hamiltonian constraint.}
\label{fig:2}
\end{figure}

\begin{figure}[t]
\begin{center}
\includegraphics[scale=0.5]{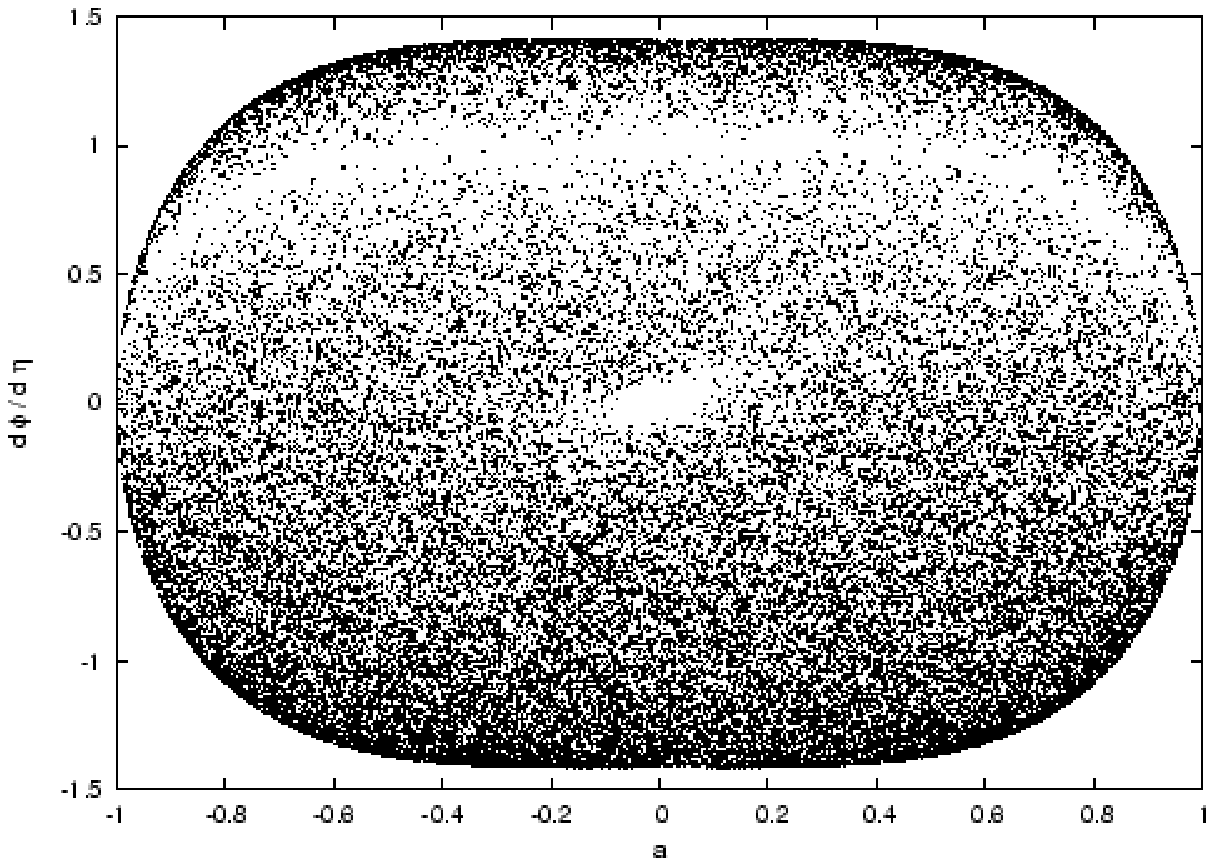}
\includegraphics[scale=0.5]{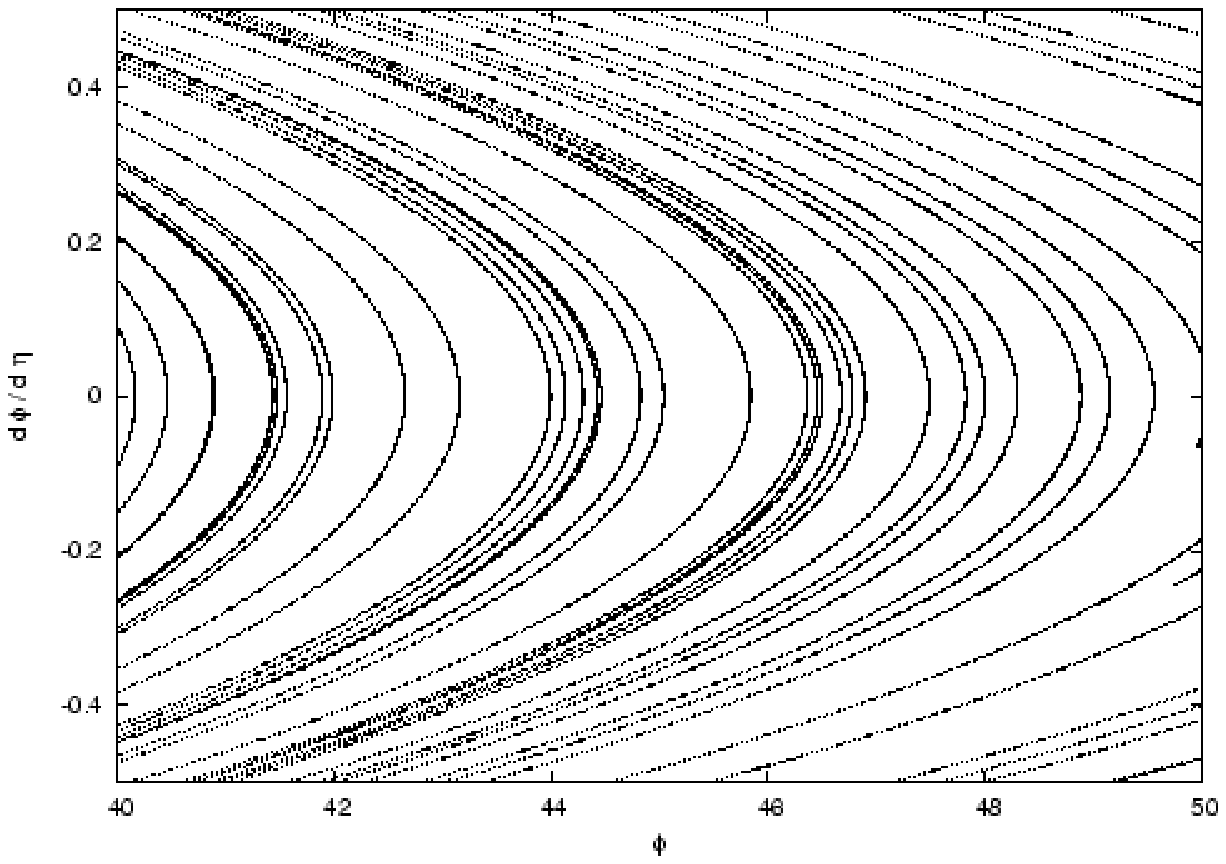}
\end{center}
\caption{The Poincar{\'e} section for the model class I, 
$a=\phi$ and $a'>0$ (upper); $a=0$, $a'>0$ (bottom). 
Motion in narrow field between two hyperbolas is completely regular. 
Complexity of behavior comes from motion near the origin of the 
configuration space.}
\label{fig:3}
\end{figure}

\begin{figure}[t]
\begin{center}
\includegraphics[scale=0.5]{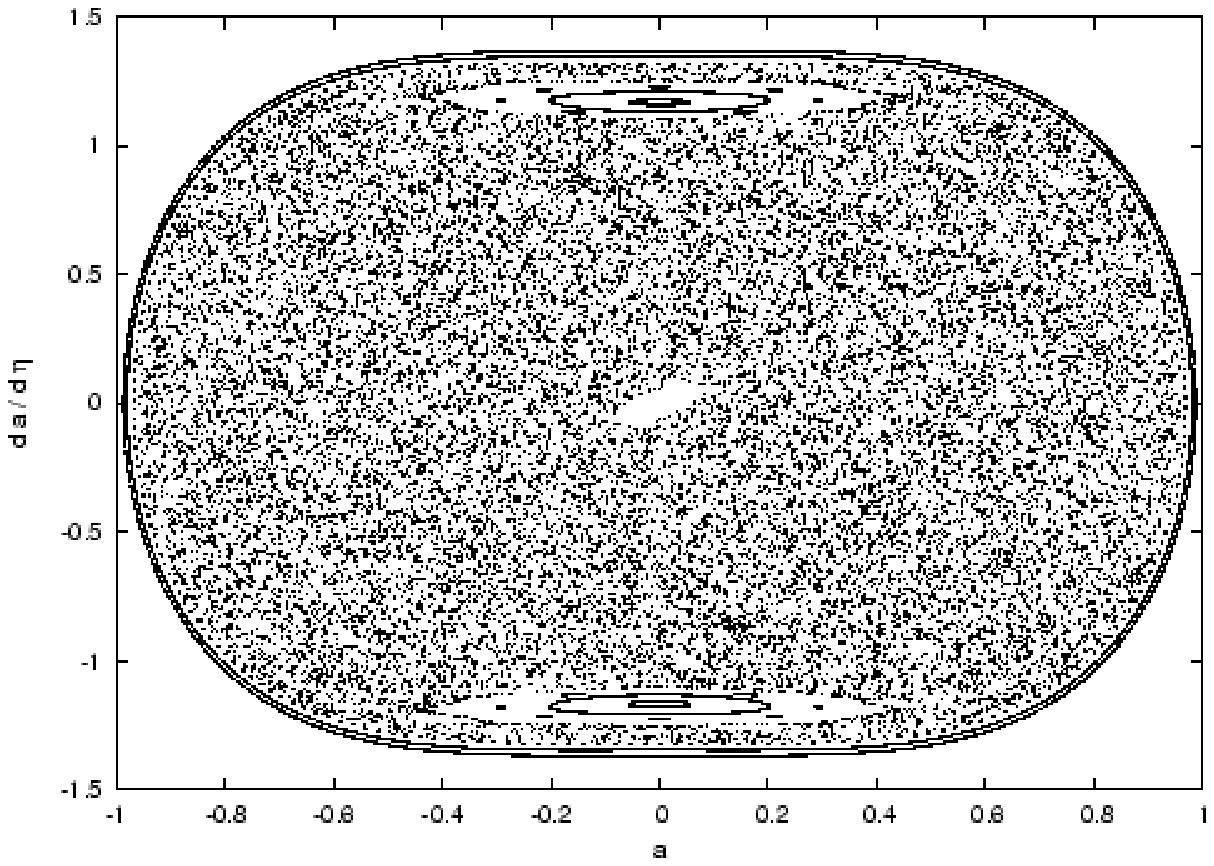}
\includegraphics[scale=0.5]{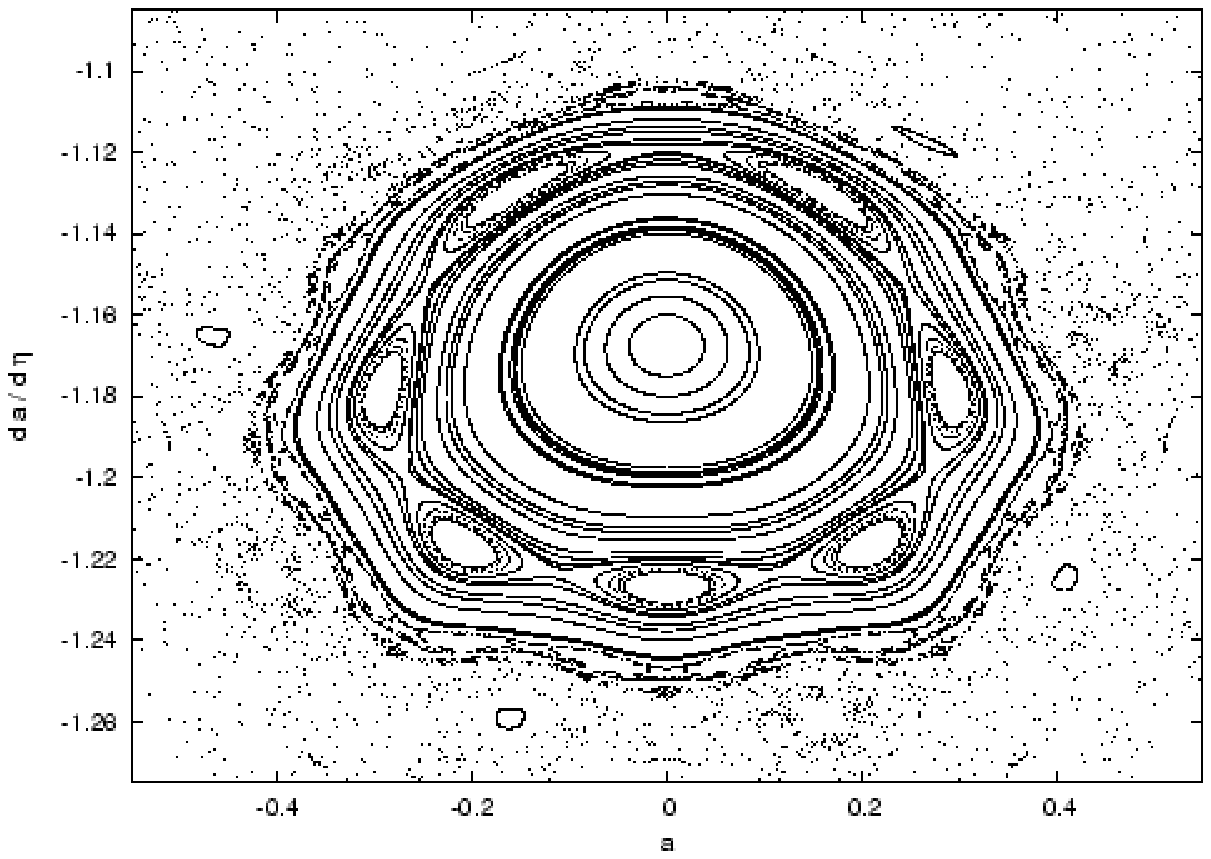}
\end{center}
\caption{The Poincar{\'e} section for the model class II, 
$\phi=0$ and $\phi'>0$. The empty region at the upper section comes 
from motion of a particle-universe along the $\phi$ ($a=0$) axis.}
\label{fig:4}
\end{figure}

\begin{figure}[t]
\begin{center}
\includegraphics[scale=1]{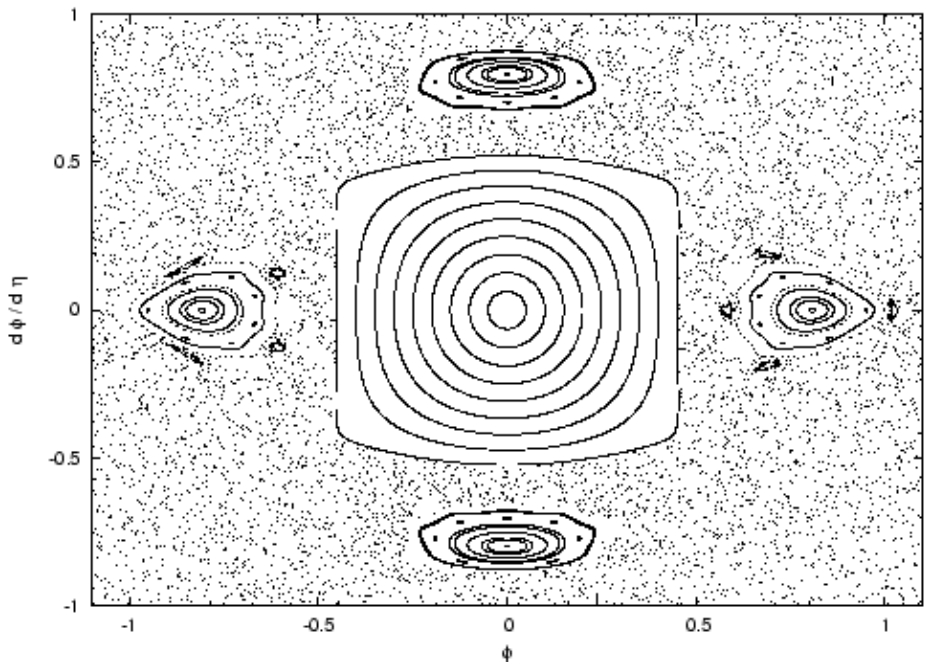}
\end{center}
\caption{The Poincar{\'e} section for the model class II, 
$a=0$ and $a'>0$.}
\label{fig:5}
\end{figure}

\begin{figure}[t]
\begin{center}
\includegraphics[scale=0.67]{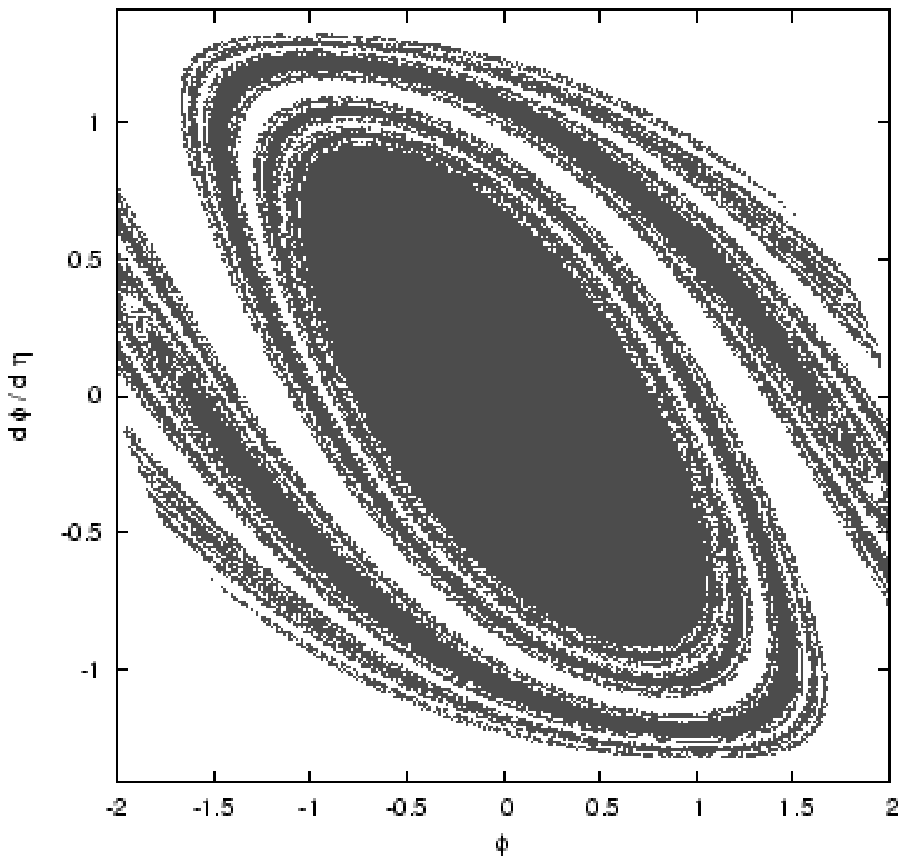} 
\includegraphics[scale=0.67]{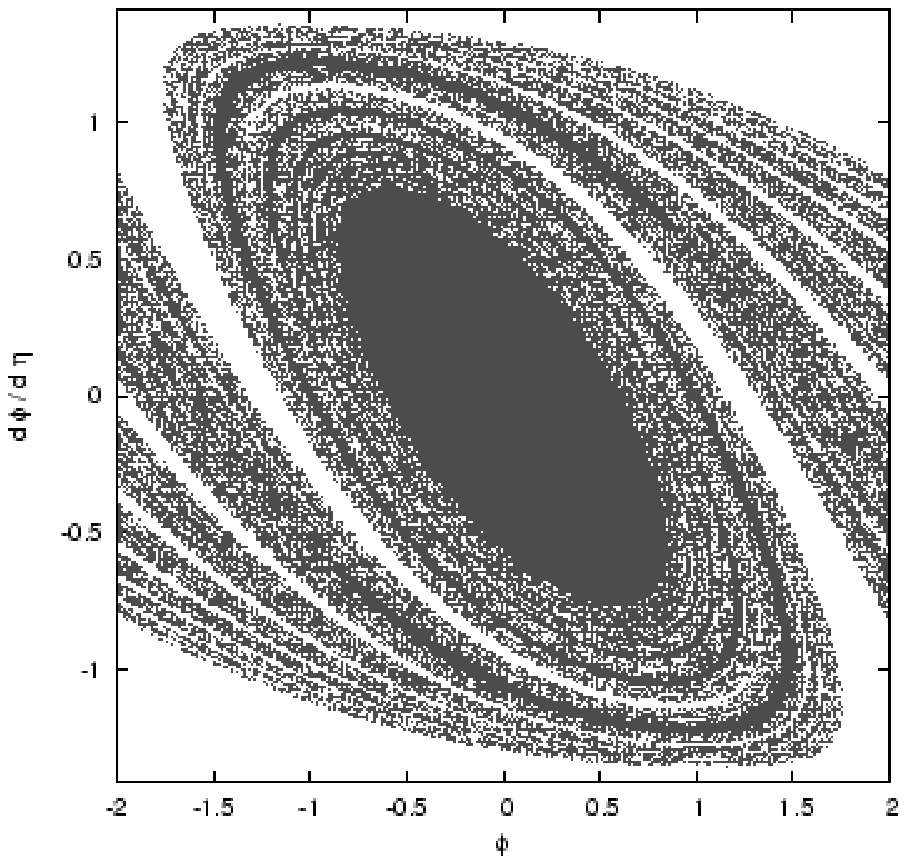}

\includegraphics[scale=0.67]{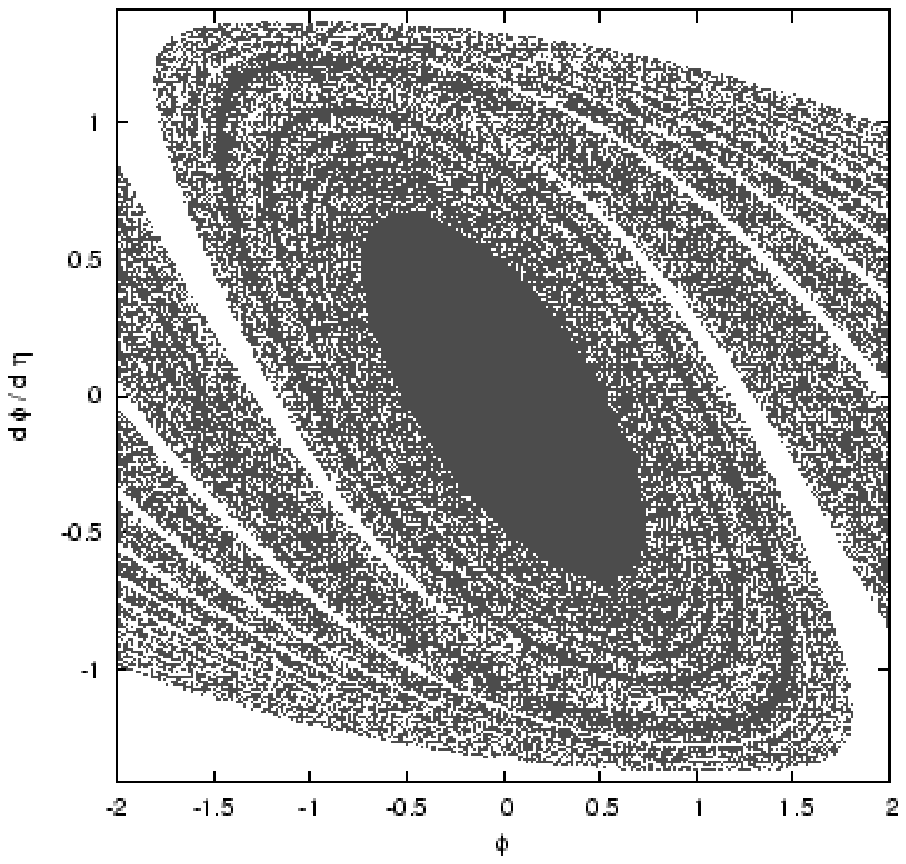}    
\includegraphics[scale=0.67]{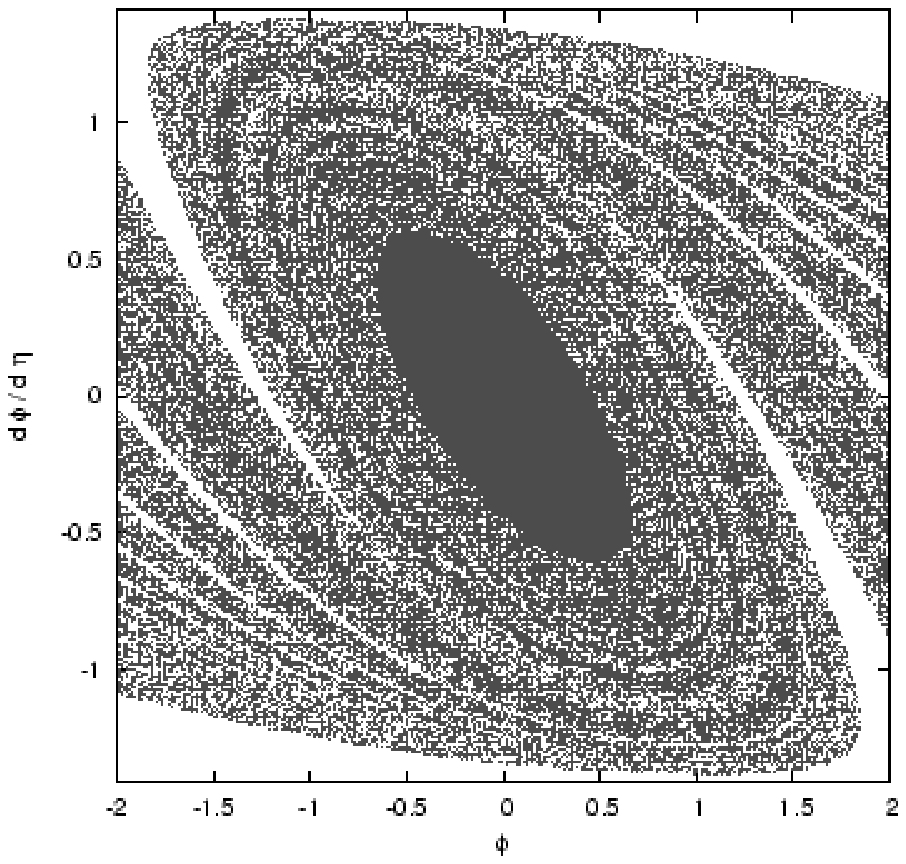}  
\end{center}
\caption{The fractal structure of the phase space of initial conditions of 
trajectories chosen at $a=0$ and $a'>0$ for class I and landing at 
$|a_{f}|=2$, $3$, $4$, $5$ (grey) or at $|\phi_{f}|=2$, $3$, $4$, $5$ (white), 
respectively. Complexity of the fractal structure increases with the final 
state of the trajectories which means that trajectories spend more time in 
the region near the point $(0,0)$ in the configuration space. The solid areas 
in the center of figures come from the motion along the $\phi$-axis.}
\label{fig:6}
\end{figure}

\begin{figure}[t]
\begin{center}
\includegraphics[scale=0.67]{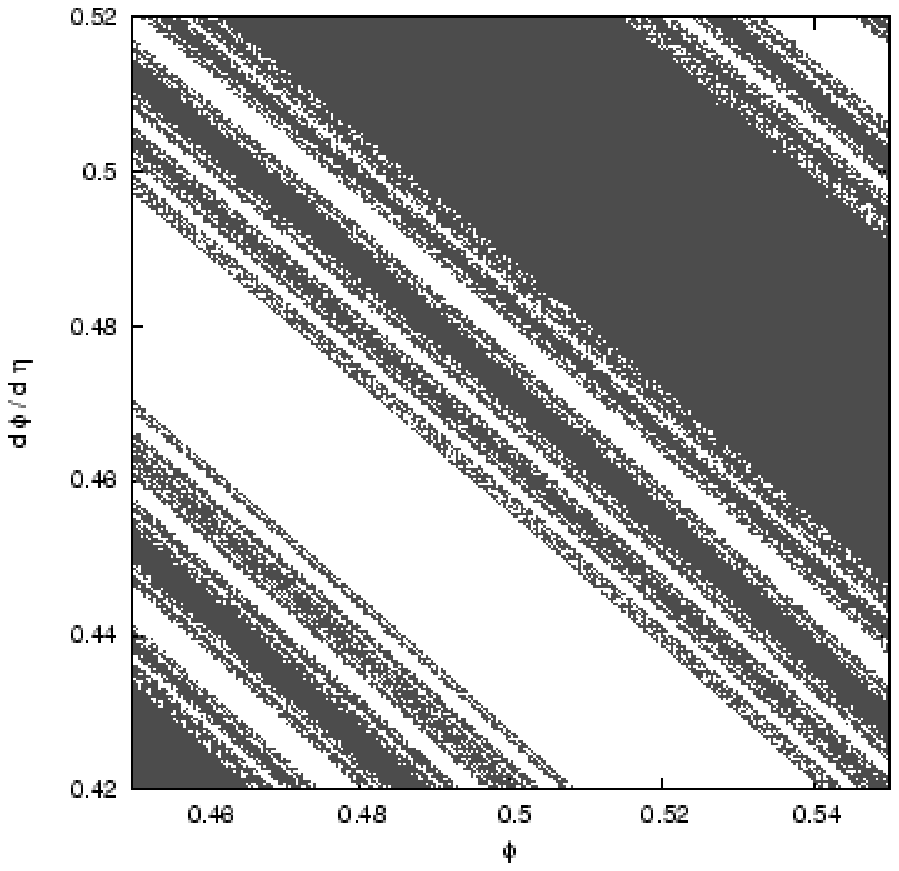}
\includegraphics[scale=0.67]{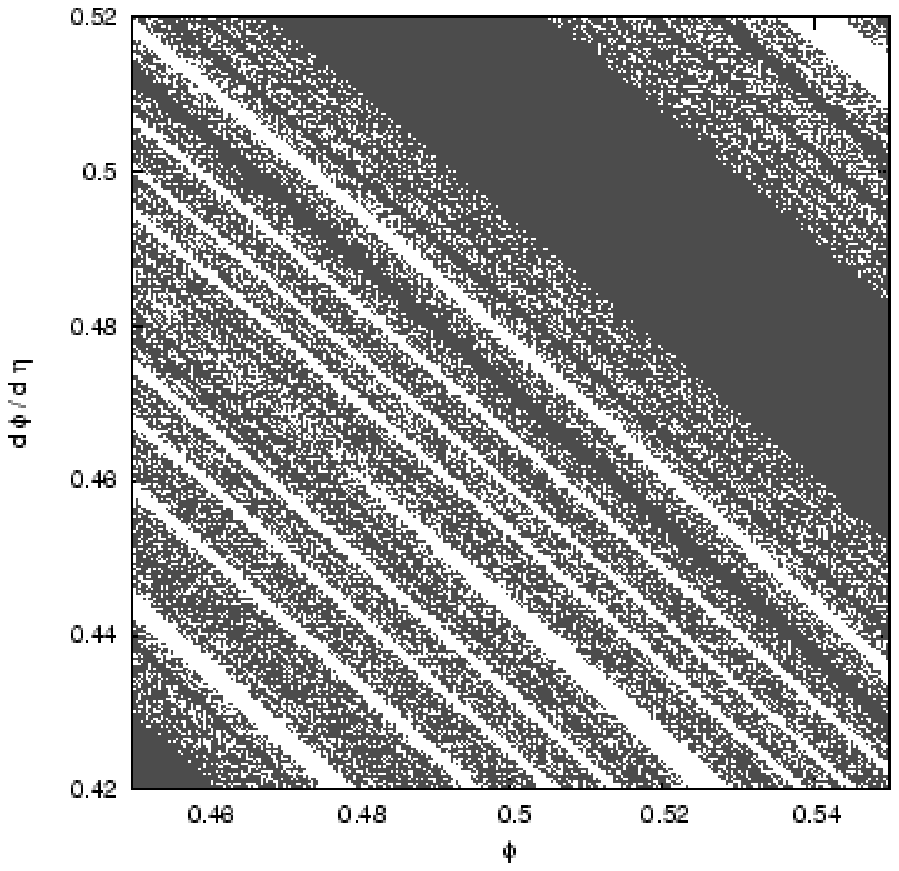}

\includegraphics[scale=0.67]{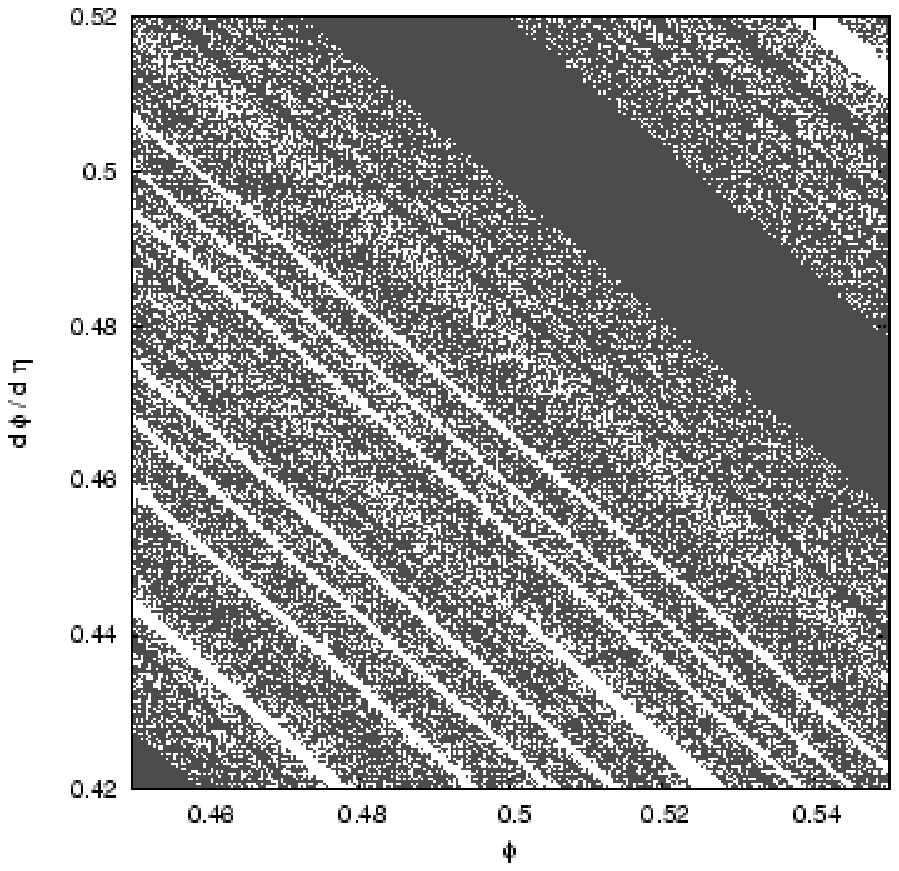}
\includegraphics[scale=0.67]{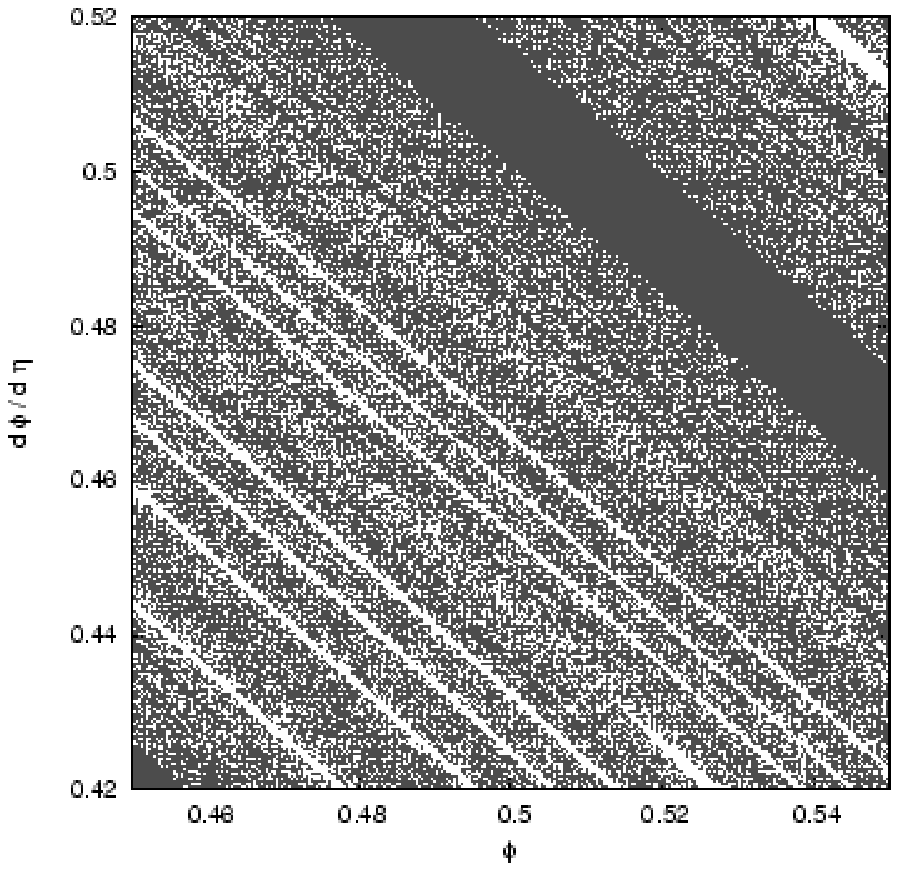}  
\end{center}
\caption{Magnification of the fractal structure of the phase space of initial 
conditions from Fig.~\ref{fig:6}.}
\label{fig:7}
\end{figure}

\begin{figure}[t]
\begin{center}
\includegraphics[scale=0.67]{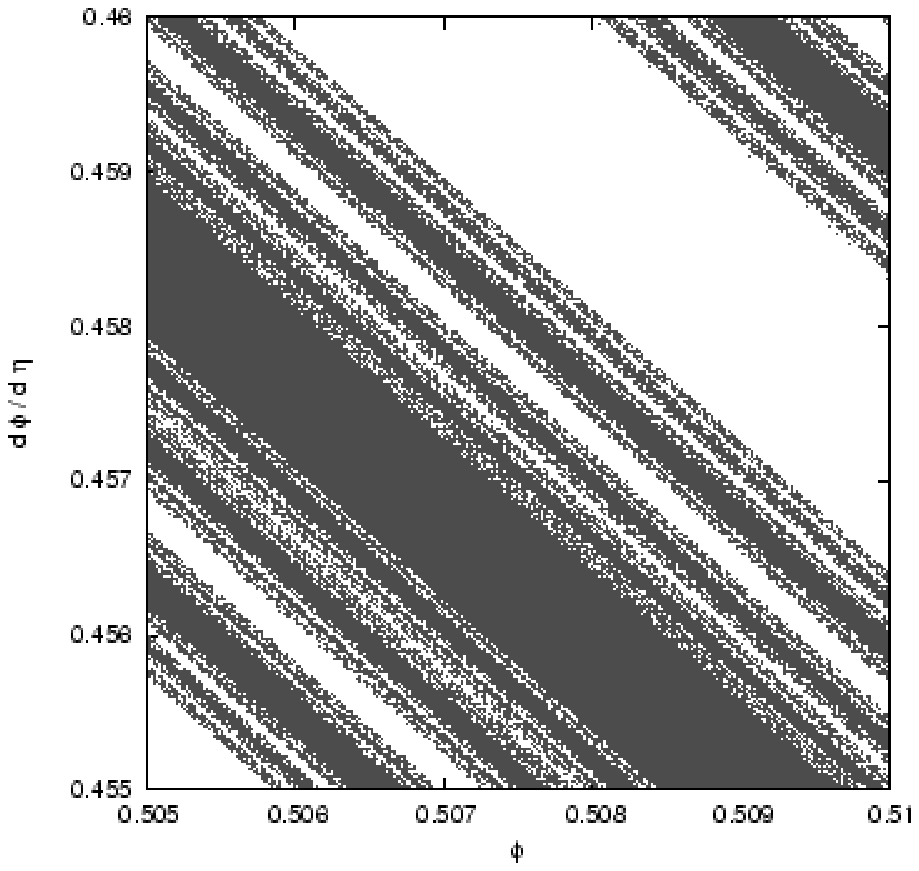}
\includegraphics[scale=0.67]{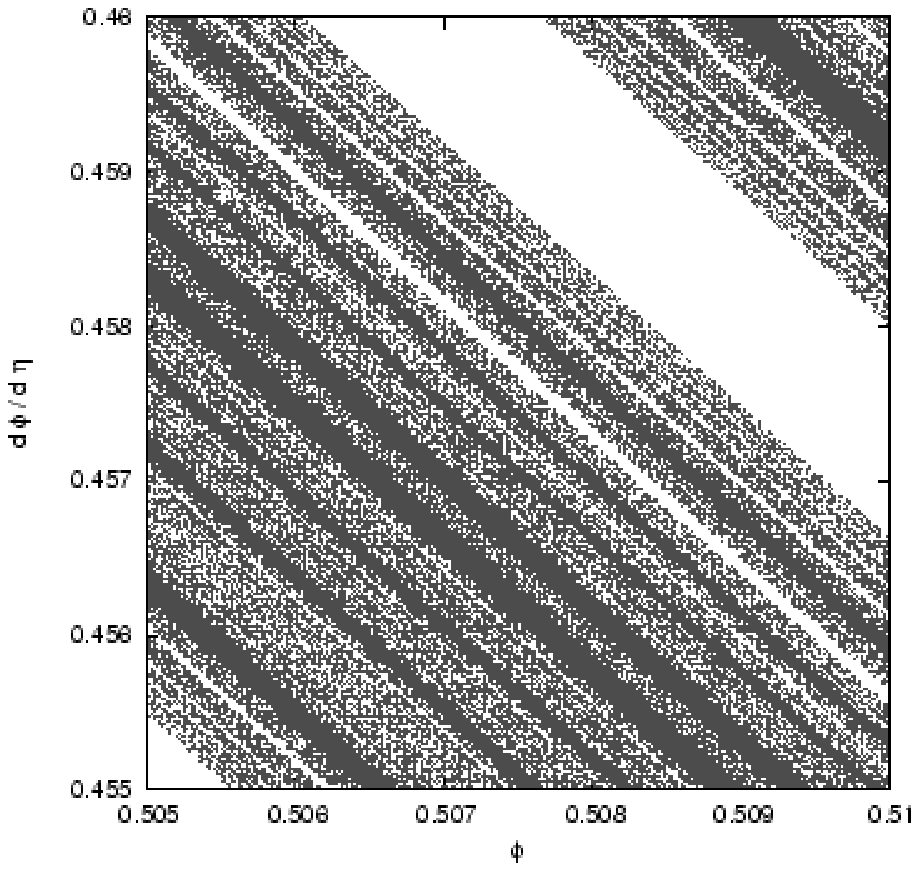}  

\includegraphics[scale=0.67]{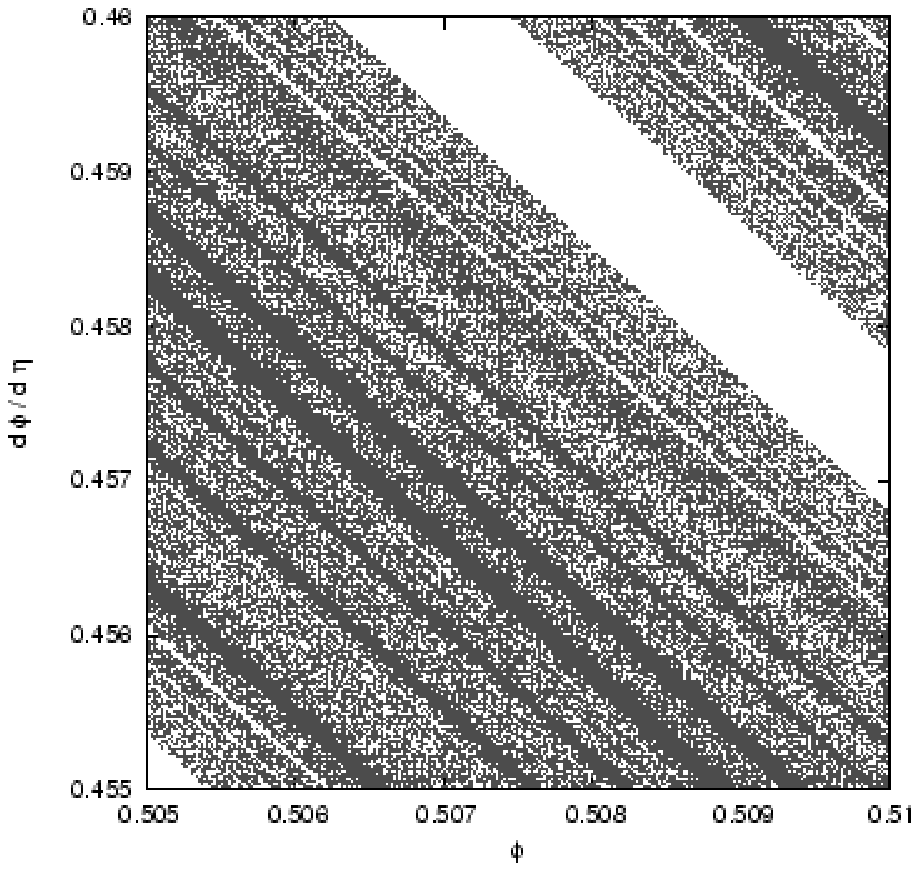}
\includegraphics[scale=0.67]{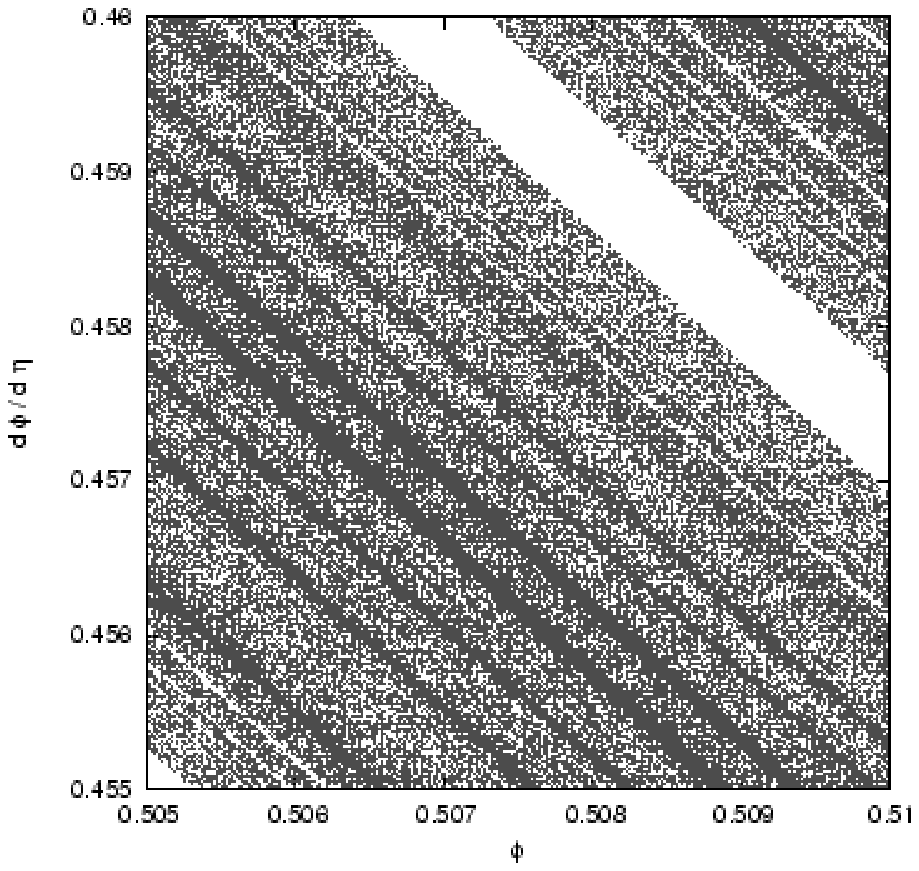}
\end{center}
\caption{Magnification of the fractal structure of the phase space of initial 
conditions from Fig.~\ref{fig:7}.}
\label{fig:8}
\end{figure}

The idea of the description of a dark energy field in terms of a Higgs field 
which creates inertial mass through the spontaneously symmetry breaking has 
been investigated lately \cite{Nemiroff:2004jf}. For $m^2 < 0$ we can 
distinguish two subclasses of the model with the spontaneously symmetry 
breaking for which chaotic behavior can be detected
\begin{eqnarray*}
\mbox{class I:} &\qquad A&=B=C=D=0, \quad E > 0 \mbox{ (or } m^2 < 0), \\
\mbox{class II:} &\qquad A&=B=D=0, \quad C>0 \mbox{ (or } \Lambda<0), 
\quad E>0 \mbox{ (or } m^2 < 0).
\end{eqnarray*}
These are flat models with conformally coupled phantom fields for which 
trajectories have the property of topological transitivity in contrast to 
the case $m^2 > 0$. In Fig.~\ref{fig:1} and \ref{fig:2} we present sample
trajectories for both models and evolution for every trajectory is last for the
same interval of time. From this simple picture we can initially conclude that
the systems are sensitive on initial conditions.

The first class of models is isomorphic with the well known Yang-Mills systems 
with the potential function $V \propto x^2 y^2$. Because the domain admissible 
for motion is bounded and its boundary has negative curvature, the property of 
recurrence of trajectories is present. We consider the both classes of systems 
on some distinguished energy level $\mathcal{H} = \mathcal{E} \propto 
\rho_{\rm{r},0} > 0$. The Poincar{\'e} sections for 
both classes are represented in Figs.~\ref{fig:3},~\ref{fig:4},~\ref{fig:5}. 
In Fig.~\ref{fig:3} it is shown the Poincar{\`e} section for the flat 
cosmological model with vanishing $\Lambda$, $\lambda$, and $m^{2} < 0$ (class I). 
If we postulate additionally the existence of radiation matter in the model 
than we deal with the system on the constant non-zero energy level. 
In this case we obtain some chaotic distribution of points on the Poincar{\`e} 
section $(a,\phi')$. In Fig.~\ref{fig:4} and ~\ref{fig:5} it is presented 
the flat cosmological model with the negative cosmological constant $\Lambda$ 
and $m^{2} < 0$ (class II) on the plane $(a,a')$. In this case we can 
observe islands of stability. However, the trajectories wander to the 
non-physical region of $a<0$. This allows many cycles to be considered if we 
continue the scale factor into negative values. But what it means physically 
is not clear.

All these figures illustrate what we proved earlier, namely the standard 
chaos with recurrence of orbits and the property of sensitive dependence 
on initial conditions. 

We cannot perform the analogous analysis for the class of models with $m^{2} > 0$ 
because there is no chaos in Wiggins' standard sense. However they possess the 
property of complex behavior of trajectories similar to the non-chaotic 
scattering process. Note that the evidence of chaos in terms of fractal basins, 
Cantori or stochastic layers requires the recurrence of trajectories. Similarly 
the Poincar{\'e} sections can be constructed from many cycles for a useful 
picture to emerge \cite{Cornish:1995mf}.

In the case of $m^{2} < 0$, trajectories of the system return to the 
neighborhood of the origin time after time and we have the multiple scattering 
process on the potential walls (boundaries of the domain admissible for 
motion). We can control and count how many times the trajectory gets closer to 
the origin than some fixed distance from the origin. With increasing this fix 
distance the number of controlled scattering events increases and a more 
complicated fractal structure arises. Fig.~\ref{fig:6}, \ref{fig:7} and 
\ref{fig:8} illustrate fractal structures of the phase space of initial 
conditions for trajectories reaching some values $a_f$ and $\phi_f$ at some 
moment of time evolution. The fractal dimension calculated by counting cells 
which contain light and dark areas (i.e. leading to two types of the final 
outcome) is $a_{f}=\phi_{f}=2$ --- $D_{0}=1.676$; $a_{f}=\phi_{f}=3$ --- 
$D_{0}=1.888$; $a_{f}=\phi_{f}=4$ --- $D_{0}=1.953$; $a_{f}=\phi_{f}=5$ --- 
$D_{0}=1.971$. The chosen method of counting cell enables us to order fractals 
with respect to the increasing complexity, i.e. for a nearly regular system the 
fractal dimension is close to one, in turn for a completely chaotic one is 
approaching two. One can also observe that complexity of dynamical behavior of 
trajectories grows more the longer trajectories come more often in the 
neighborhood of the origin of the coordinate system because the motion 
between branches of hyperbolas is regular (see Fig.~\ref{fig:3}).

\begin{figure}[t]
\begin{center}
\includegraphics[scale=0.67]{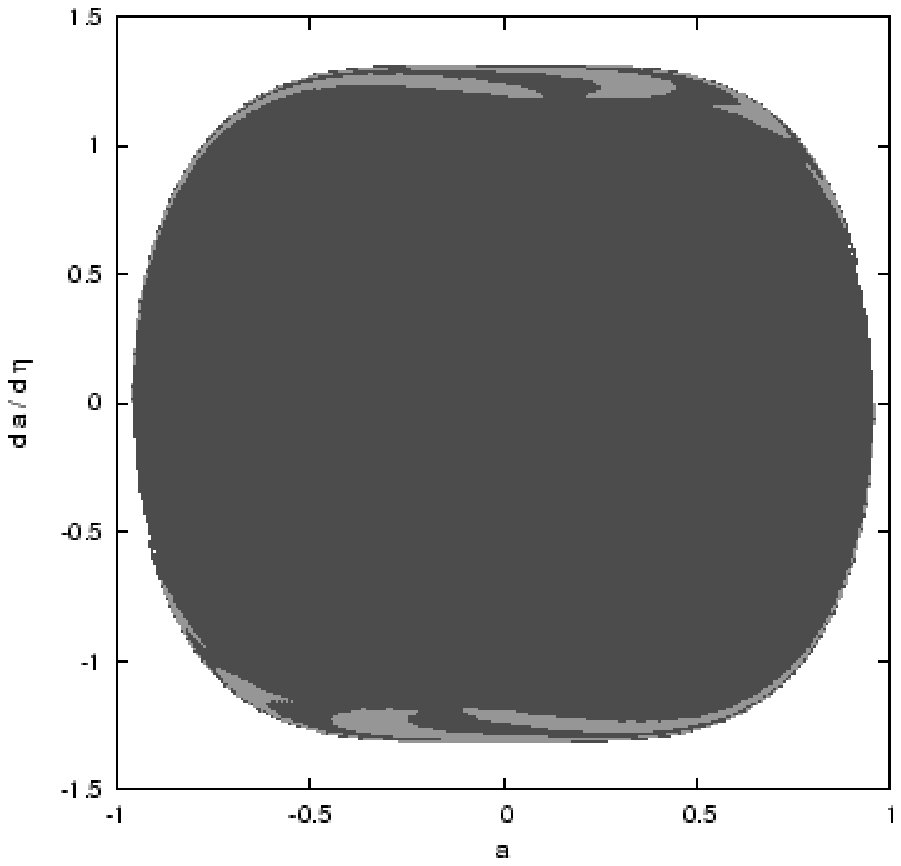}
\includegraphics[scale=0.67]{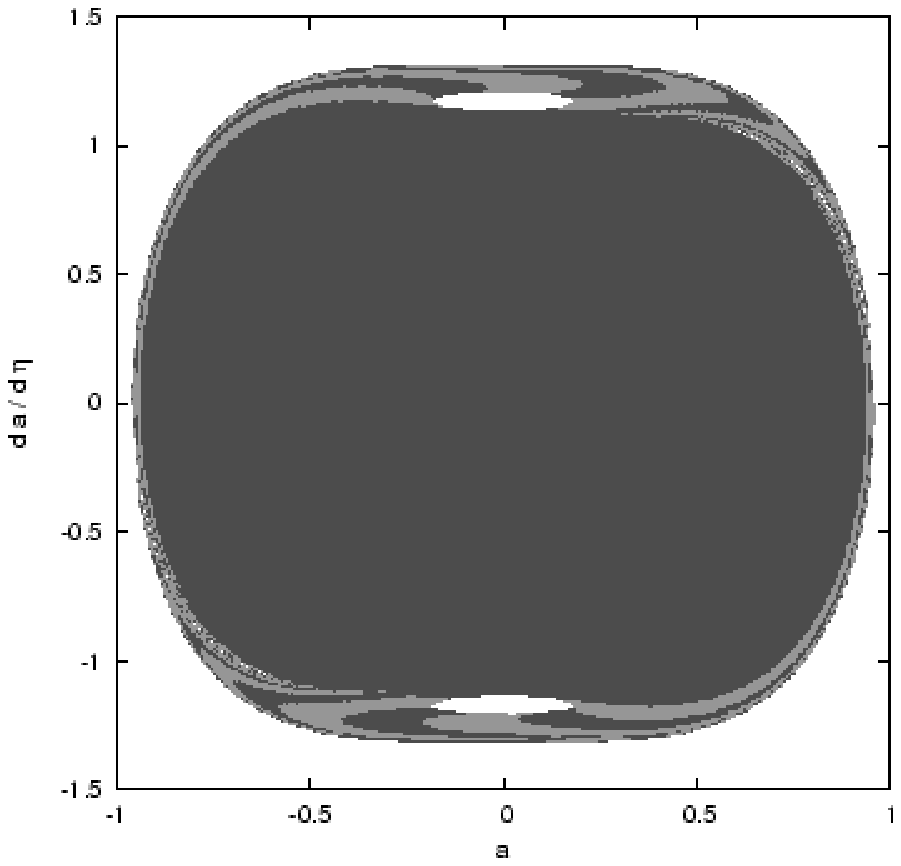}
\end{center}
\caption{The structure of phase space of initial conditions for class II 
trajectories chosen at $\phi=0$ and $\phi'>0$ and landing at $\phi=0.75$ 
(dark grey) and $\phi=-0.75$ (grey) presented in left figure and $\phi=0.9$ 
(dark grey) and $\phi=-0.9$ (grey) in right figure. The white poles correspond 
to the centers of stability in Fig.~\ref{fig:4}.}
\label{fig:9}
\end{figure}

\begin{figure}[t]
\begin{center}
\includegraphics[scale=0.67]{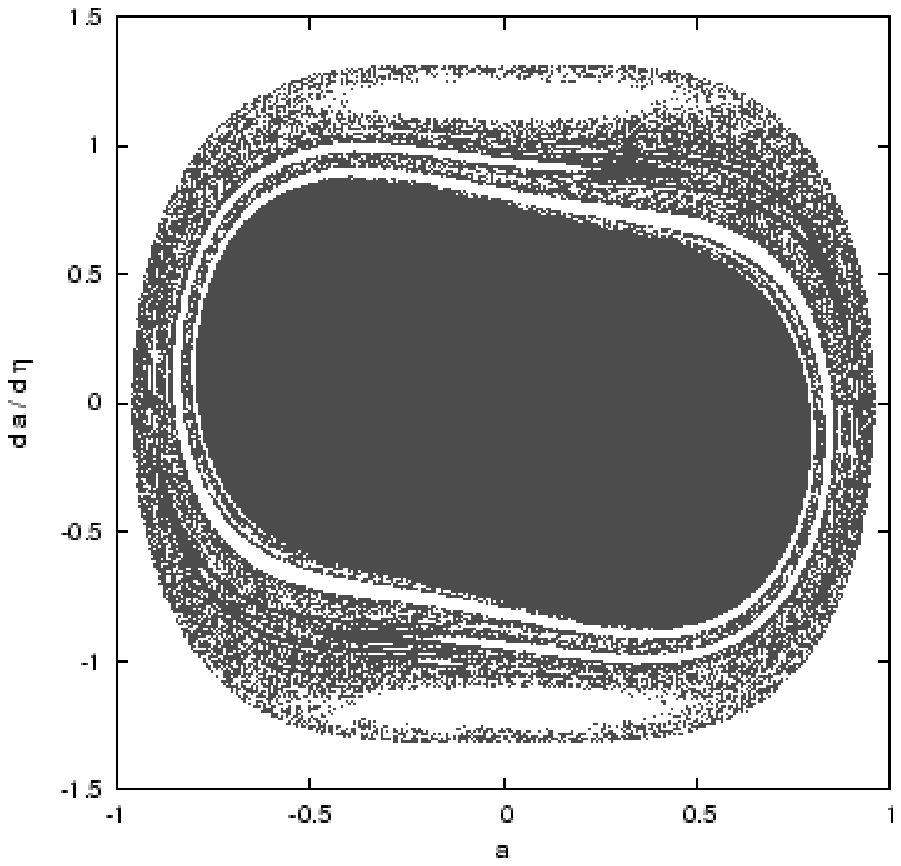}
\includegraphics[scale=0.67]{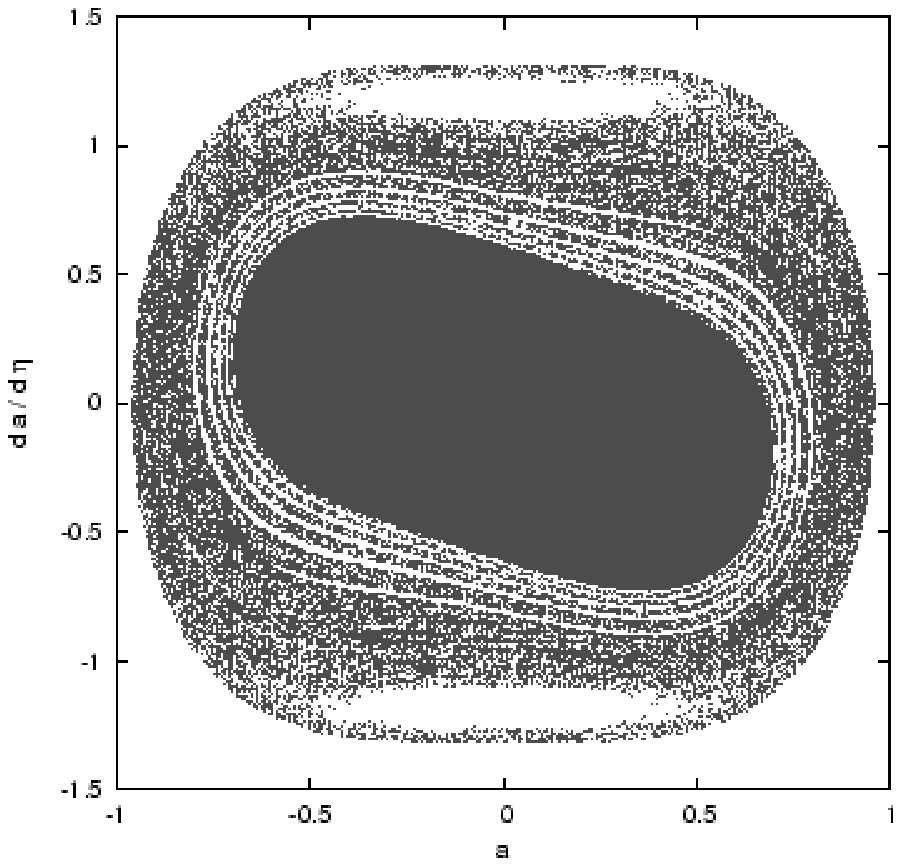}
\end{center}
\caption{The fractal structure of phase space of initial conditions of 
trajectories class II landing at $\phi=2$ (left) and $\phi=3$ (right). 
Initial conditions leading to negative values of $\phi$ omitted for clarity.}
\label{fig:10}
\end{figure}

\begin{figure}[t]
\begin{center}
\includegraphics[scale=0.67]{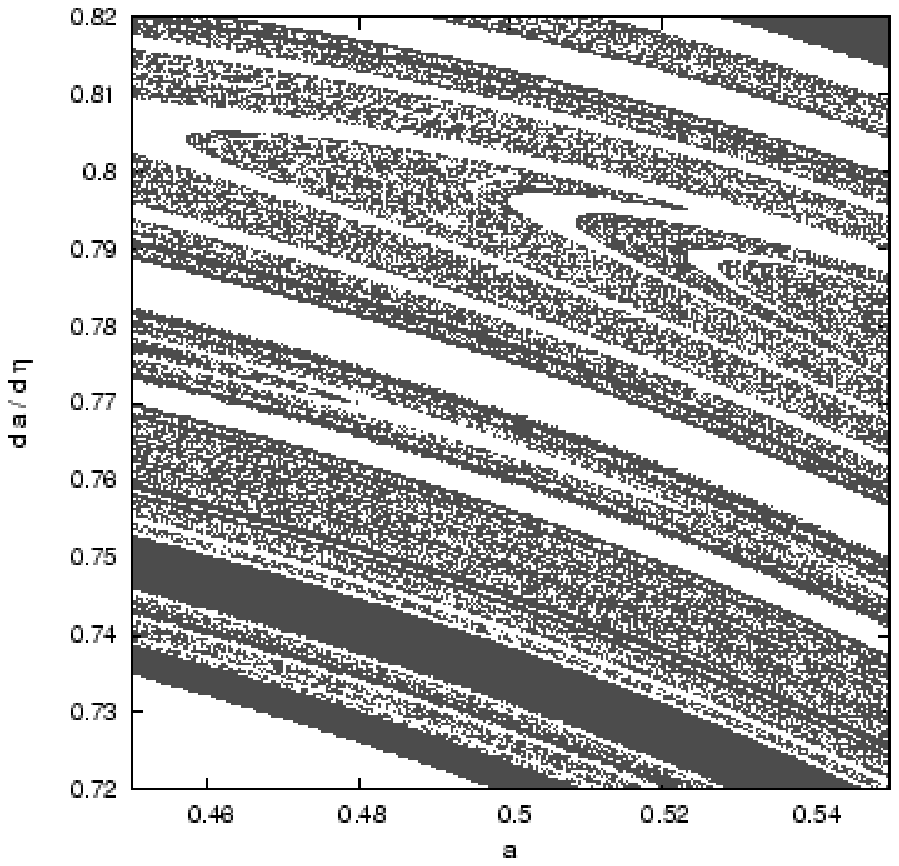}
\includegraphics[scale=0.67]{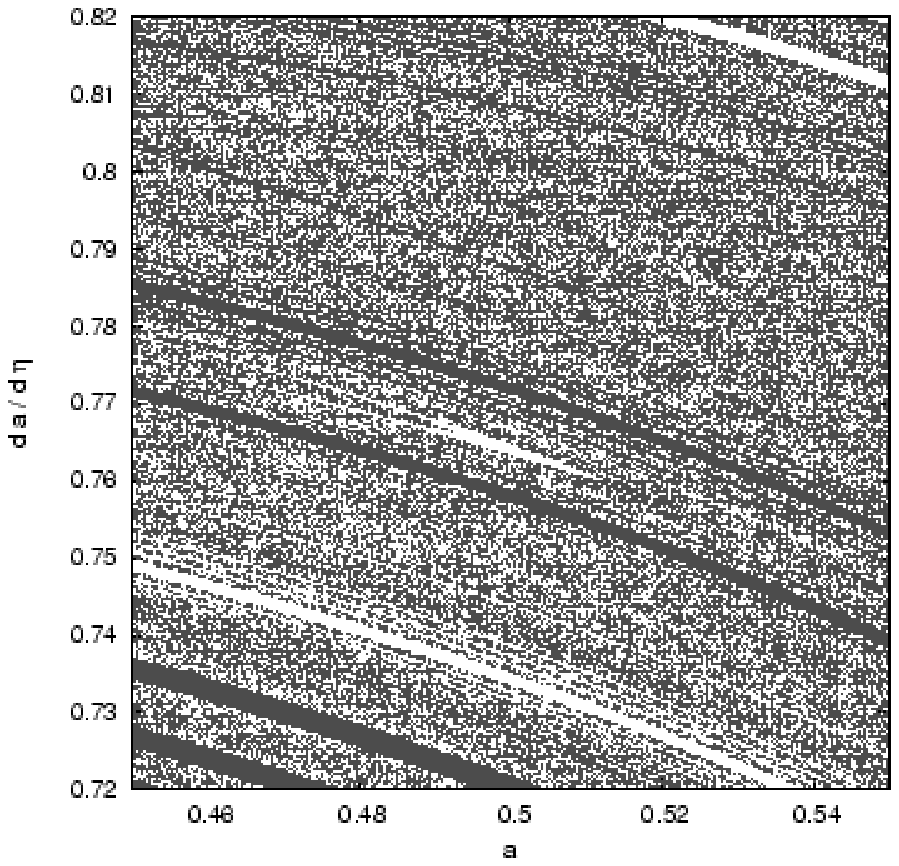}
\end{center}
\caption{Magnification of the fractal structure of the phase space of initial
conditions of trajectories class II from Fig.~\ref{fig:10}.}
\label{fig:11}
\end{figure}

\begin{figure}[t] 
\begin{center}
\includegraphics[scale=0.67]{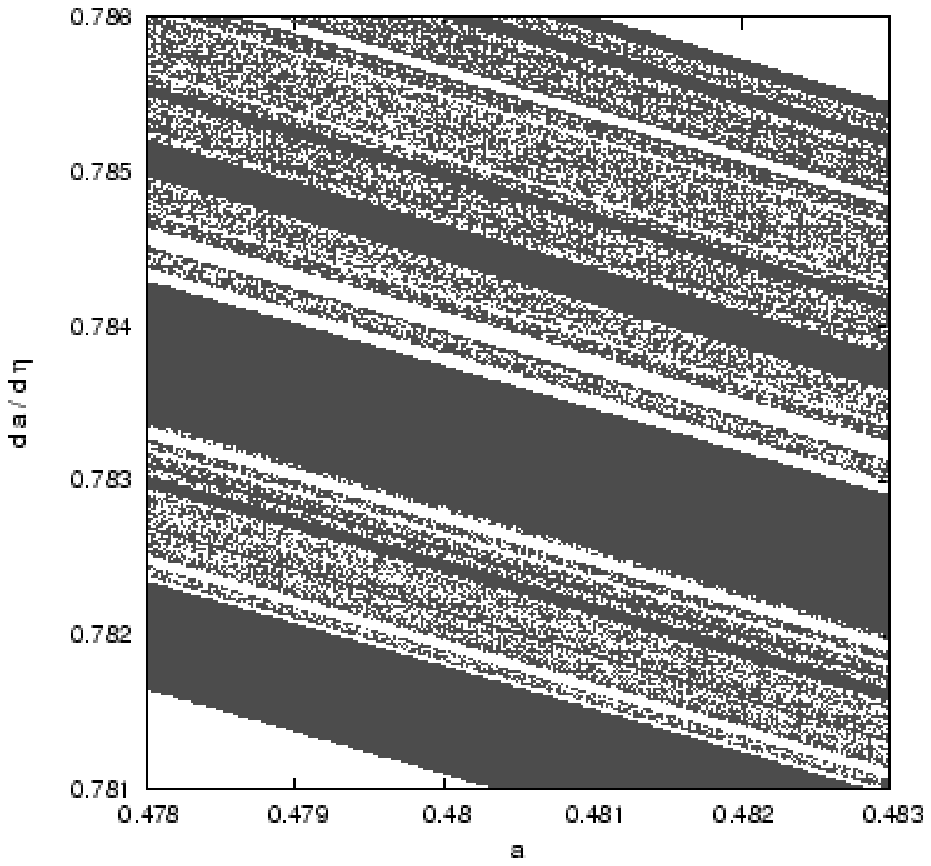}
\includegraphics[scale=0.67]{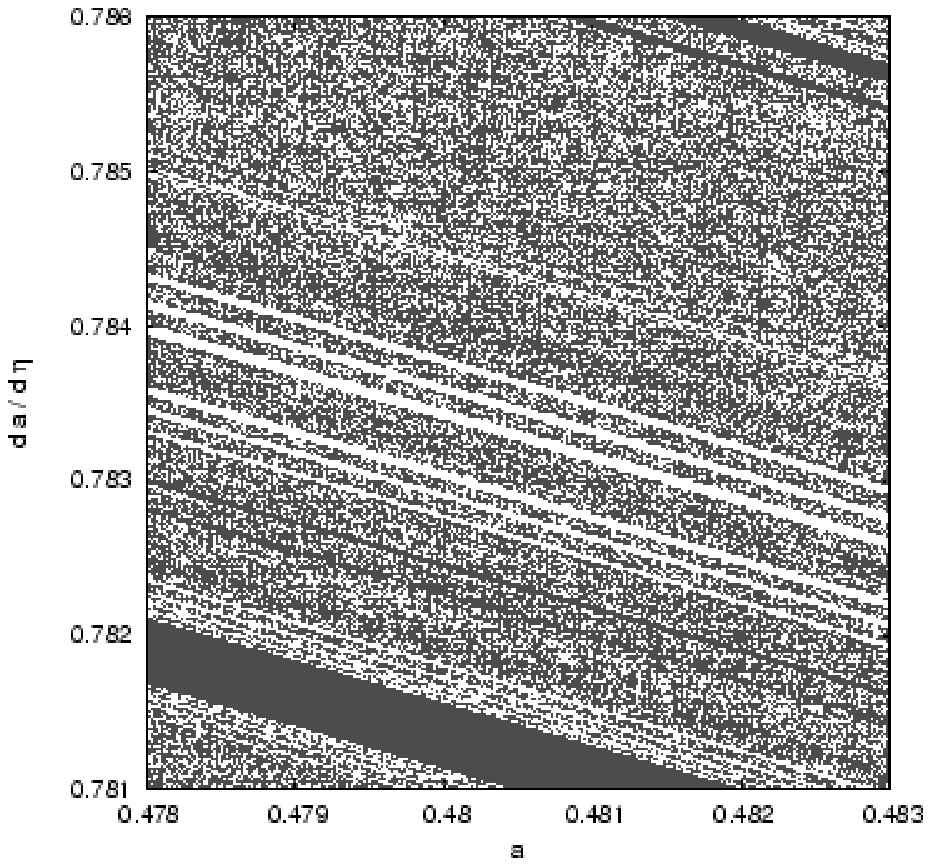}
\end{center}
\caption{Magnification of the fractal structure of the phase space of initial
conditions of trajectories class II from Fig.~\ref{fig:11}.}
\label{fig:12}
\end{figure}

Fig.~\ref{fig:9},~\ref{fig:10},~\ref{fig:11} and \ref{fig:12} present the 
fractal structure of phase space of initial conditions for the category II model. 
At enough small value of $\phi_f$ the system is regular and no chaotic 
behavior is present (Fig.~\ref{fig:9}). With an increasing value of $\phi_f$:
$\phi_{f} = 2$ --- $D_{0} = 1.812$; $\phi_{f} = 3$ --- $D_{0} = 1.967$,
we observe the transition to chaos which cause to emerge the fractal structure 
of the space of initial conditions (Fig.~\ref{fig:10}--\ref{fig:12}).

\begin{figure}[t] 
\begin{center}
\includegraphics[scale=0.67]{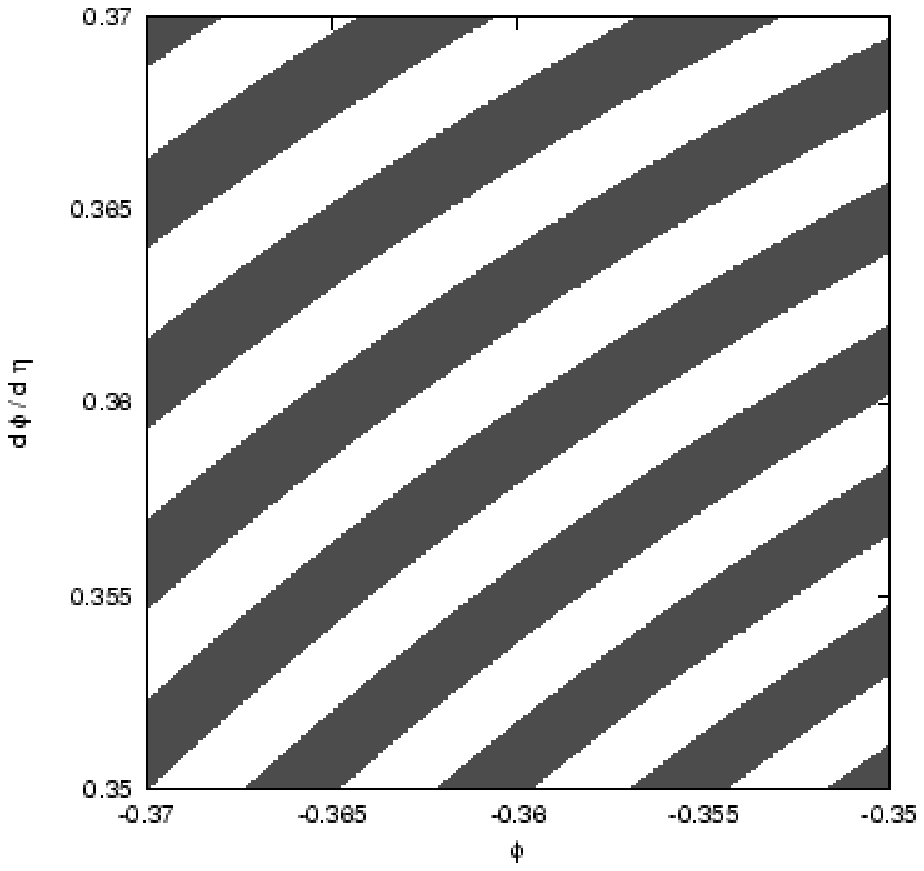}
\includegraphics[scale=0.67]{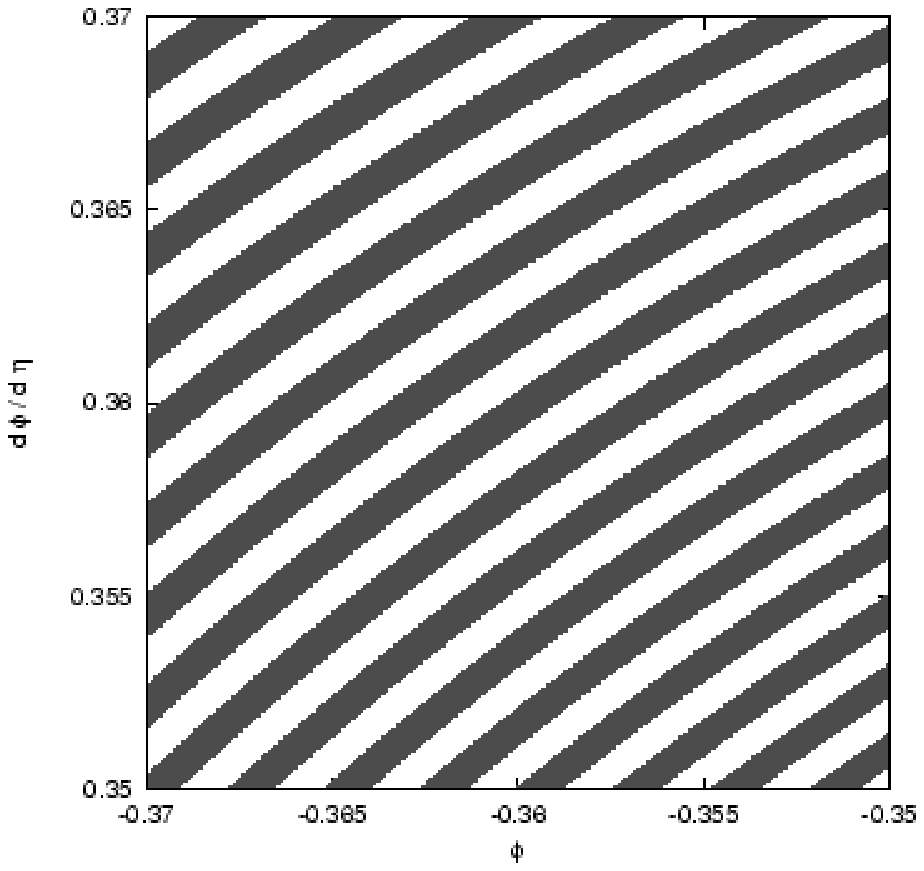}
\end{center}
\caption{The structure of phase space of initial conditions for model class I 
chosen at $a=0$ with $a'>0$ for trajectories leading to $a'=0$ (or 
equivalent $H=0$) with $\phi>0$ (grey) and $\phi<0$ (white) -- the left image, 
and initial conditions of trajectories leading to $a=0$ with $\phi>0$ (grey) 
and $\phi<0$ (white) -- the right image. There is no chaos in one cycle of 
evolution.}
\label{fig:13}
\end{figure}

In Fig.~\ref{fig:13} we plot the phase space of initial conditions chosen at
$a=0$ and $a'>0$ calculated from the Hamiltonian which lead to a given property 
in one cycle of evolution, i.e. to maximal expansion with a positive value of 
$\phi$ (grey) or negative value of $\phi$ (white) -- left image, and back to a 
final singularity again with a positive value of $\phi$ (grey) or a negative
value of $\phi$ (white) -- right image. Therefore if we consider evolution in 
a physical domain between initial and final singularities this system cannot 
be chaotic. If we prolong its evolution to the non-physical domain $a<0$ 
then we obtain chaos. Note that non-integrability indicators measure the true 
intrinsic complexity of the system in both cases \cite{Szydlowski:2004jv}.

\section{Phantom cosmology as the scattering process}
\label{sec:4}

In this section we investigate behavior of a model without the spontaneously 
symmetry breaking. This is a model with $A=B=C=D=0$ and $E=-1$ ($m^2=2$) on 
different energy levels $\mathcal{E}>0$, $\mathcal{E}=0$ and $\mathcal{E}<0$.

In the analysis of this case we explore analogy to the classical system in 
which appears the chaotic scattering. From the equation of motion we obtain 
that for any initial condition trajectories escape to infinity ($\phi \to \pm 
\infty$ $a \to \pm \infty$ as time goes to infinity). For energy 
$\mathcal{E} > 0$ the configuration space is unbounded and trajectories pass 
from one quadrant to another. On the other hand for $\mathcal{E} \le 0$ 
motion is restricted to only this quadrant of the configuration space where 
the initial conditions are. 

\begin{figure}[t]
\begin{center}
\includegraphics[scale=0.75]{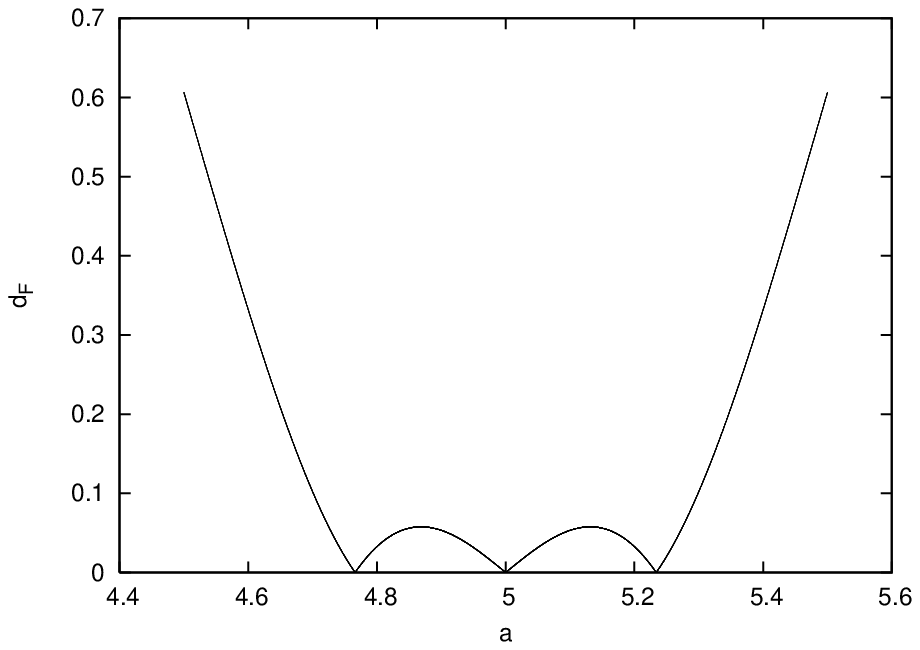} \\
\includegraphics[scale=0.75]{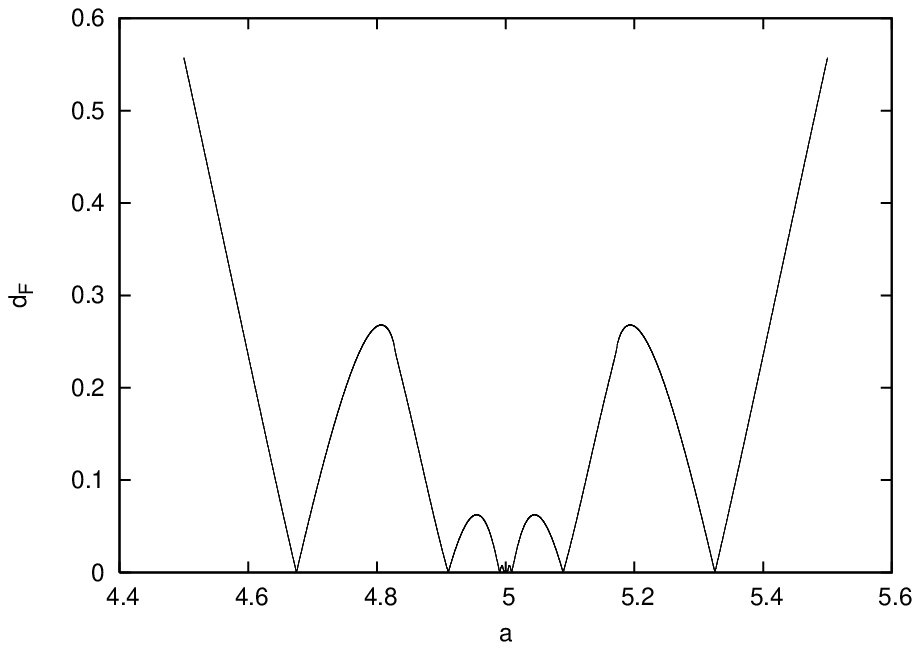} \\
\includegraphics[scale=0.75]{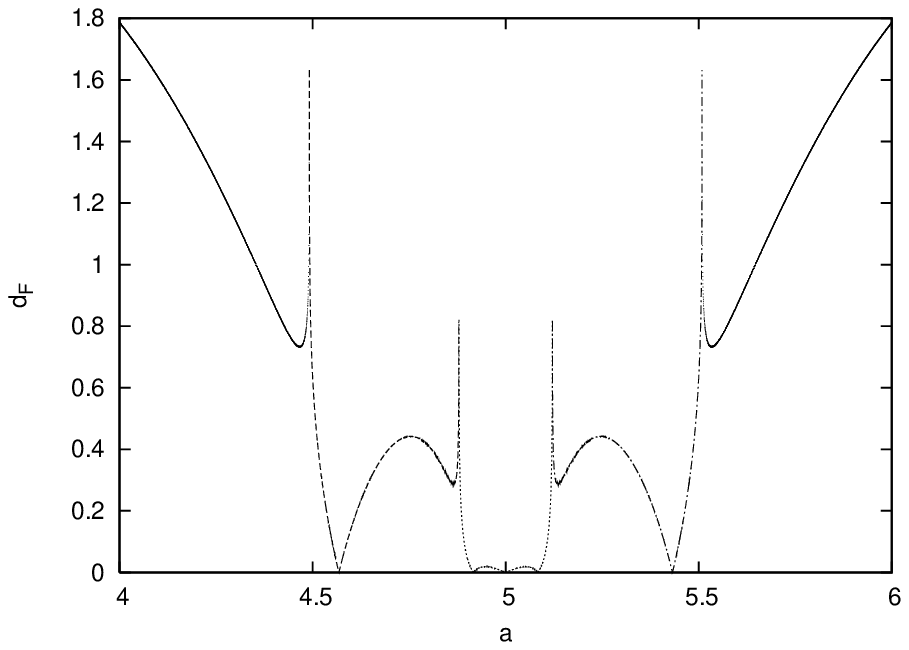}
\end{center}
\caption{The scattering process in the class A model without the spontaneously 
symmetry breaking. For all trajectories initial conditions were chosen at line
$\phi=-a+10$ in configuration space and $\dot{a}=\dot{\phi}<0$ calculated
from the Hamiltonian. On the $x$-axis we put initial position $a$ and on the 
$y$-axis we put a final distance from a symmetry point in the configuration 
space $(5,5)$ when a trajectory is escaping to infinity. For $\mathcal{E}<0$ 
(a) and $\mathcal{E}=0$ (b) all the trajectories escape to infinity 
($a \to \infty, \phi \to \infty$) without crossing $a=0$ or $\phi=0$. For 
$\mathcal{E}>0$ (c) some trajectories change a quarter after starting from 
I quarter: I -- solid line, II -- dashed line, III -- dotted line and IV -- 
dash-dot line.}
\label{fig:14}
\end{figure}

In Fig.~\ref{fig:14} we present the analysis of dependence of scattering on 
the energy level. For this aim the initial condition in the configuration space 
are chosen on the line $\phi = -a +10$ with $a' = \phi' < 0$ under 
the conservation of the Hamiltonian constraint. For $\mathcal{E} \le 0$ 
trajectories go toward the origin of coordinate system and after some time 
escape to infinity passing through a line $\phi = -a +10$. On the $x$-axis we 
mark the initial value of $a$ and $y$-axis we mark final distance from the 
symmetry point with coordinates $(5,5)$ at the moment of intersection of line 
of initial conditions. In this figure there is no discontinuities, we do not 
also observe the fluctuation of distance $d_{f}$. It enable us to conclude that 
we deal with the scattering process although the domain of interaction is not 
finite \cite{Ott:1993cd}. So there is no chaotic scattering. The motion of the 
system is regular and there is no sensitivity of motion with respect to initial 
conditions. 

\begin{figure}[t]
\begin{center}
\includegraphics[scale=0.75]{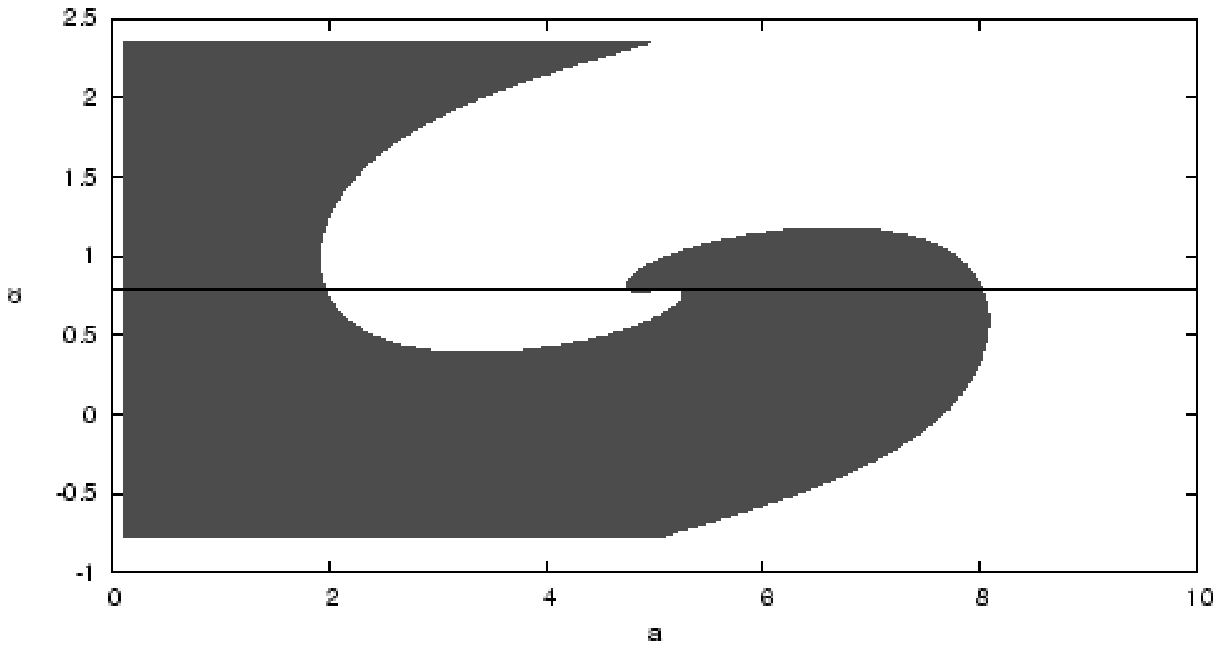} \\
\includegraphics[scale=0.75]{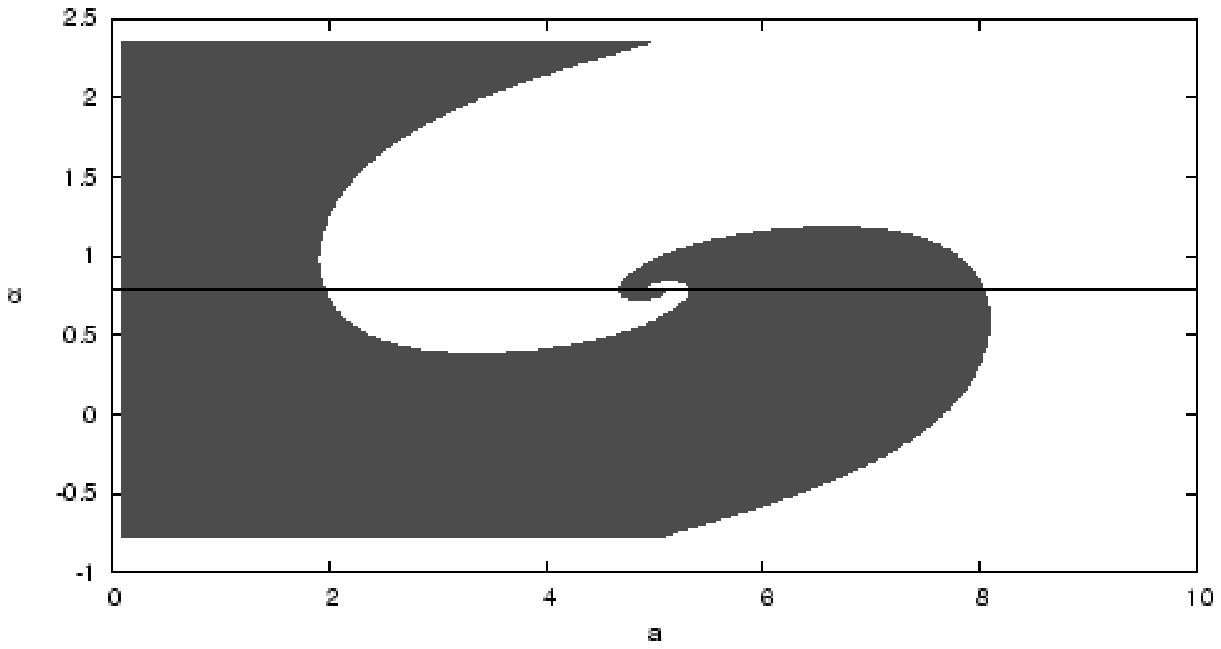} \\
\includegraphics[scale=0.75]{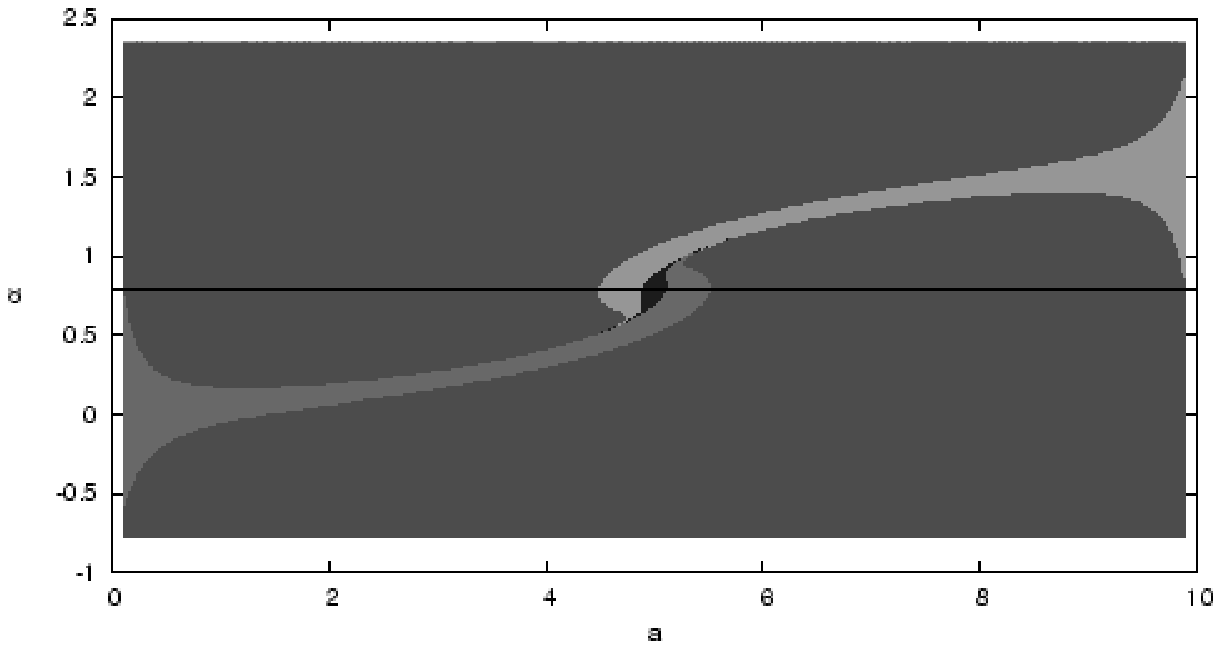}
\end{center}
\caption{The scattering process in the class A model without the spontaneously 
symmetry breaking for a large number of trajectories. (See text for 
description). Fig.~(a) and (b) correspond to $\mathcal{E}<0$ and 
$\mathcal{E}=0$, respectively. The grey areas correspond to initial conditions 
for which trajectories pass through the line $\phi=-a+10$ with $\phi>a$. 
Fig.~(c) corresponds $\mathcal{E}>0$ (the configuration space is unbounded) 
and illustrates domains of initial conditions for which trajectories outcomes 
reach different quadrants: I quadrant -- dark grey, II quadrant -- grey,
III quadrant -- black, IV quadrant --- medium grey. The black lines represent
initial conditions from Fig.~\ref{fig:14}.}
\label{fig:15}
\end{figure}

For a deeper confirmation of this statement we study numerically a larger 
number of initial conditions (Fig.~\ref{fig:15}). In the configuration space 
the initial conditions are chosen as previously, i.e. $\phi = -a +10$ 
and the initial conditions for velocities are parameterized in a natural way 
by an angle $\alpha$, i.e. $a' = - V(a,\phi,\mathcal{E}) \cos\alpha$, 
$\phi' = - V(a,\phi,\mathcal{E}) \sin\alpha$
where $V(a,\phi,\mathcal{E}) = \sqrt{\mathcal{E} + m^{2} a^{2} \phi^{2}}$. 
In Fig.~\ref{fig:15} (a) and (b) it is presented results of our analysis for 
$\mathcal{E} < 0$ and $\mathcal{E} = 0$ respectively. The grey area 
corresponds to initial conditions for which trajectories pass through line 
$\phi = -a +10$ with $\phi > a$. Fig.~\ref{fig:15} (c) illustrates domains 
of initial conditions for $\mathcal{E} > 0$ for which trajectories outcomes 
reach different quadrants : I quadrant -- dark grey, II quadrant -- grey, 
III quadrant -- black, IV quadrant --- medium grey. The black line presents 
initial conditions taken from Fig.~\ref{fig:14}. We cannot observe chaotic 
scattering in this case. 

\section{Acceleration in Phantom Cosmology}
\label{sec:5}

The current Universe is in an accelerating phase of expansion that's why we 
check whether the scalar fields in the FRW cosmology can explain this 
phenomenon. For this aim let's are consider acceleration equation which can be 
obtained from the Raychaudhury equation
\begin{equation}
\frac{\ddot{a}}{a} = - (\rho+3 p) = -(-2\dot{\psi}^{2} - 2U(\psi)
+3 \xi H (\psi^{2})\dot{ } + 3 \xi (\psi^{2})\ddot{ } + 6 \xi \dot{H} \psi^{2} +
6 \xi H^{2} \psi^{2}) - 2\rho_{r}
\label{eq:24}
\end{equation}
In the special case of the phantom field minimally coupled to gravity $\xi=0$
and without radiation $\rho_{r}=0$ we have
\begin{equation}
\frac{\ddot{a}}{a} = 2 (\dot{\psi}^{2} + U(\psi)).
\label{eq:25}
\end{equation}
Next, for zero energy level and $\Lambda=0$ from (\ref{fint}) we have that
\begin{equation}
H^{2} = 2\rho_{\psi} = -\dot{\psi}^{2} + 2 U(\psi).
\label{eq:26}
\end{equation}
and finally we receive the acceleration equation which independent of the form 
of the potential function
\begin{equation}
\frac{\ddot{a}}{a} = 3 \dot{\psi}^{2} + H^{2} \ge 0.
\end{equation}

In the case of conformally coupled phantom field $\xi=1/6$ we have
\begin{equation}
\frac{\ddot{a}}{a} = - (-\dot{\psi}^{2} - 2U(\psi) - H^{2}\psi^{2} -
H(\psi^{2})\dot{ } + \psi \frac{\rmd U}{\rmd \psi}(\psi)) - 2\rho_{r}
\label{eq:28}
\end{equation}
and after the rescaling time and field variables: $\rmd t = a \rmd \eta$, 
$\phi=a \psi$ and inserting the potential function (\ref{eq:5}) we have
\begin{equation}
\frac{\ddot{a}}{a} = \frac{1}{a^{4}}(\phi'^{2}- \frac{1}{2}\lambda \phi^{4} -
2\rho_{r,0}),
\label{eq:29}
\end{equation}
where prime denotes differentiation with respect to the conformal time and dot 
with respect to the cosmological time. 
We can also express equation of state parameter (\ref{eq:8}) in these new
variables
\begin{equation}
w_{\phi} \equiv \frac{p_{\phi}}{\rho_{\phi}} =
\frac{\frac{1}{a^{4}}(-\frac{1}{6}\phi'^{2} - \frac{1}{6}m^{2}a^{2}\phi^{2} +
\frac{1}{12}\lambda\phi^{4})}{\frac{1}{a^{4}}(-\frac{1}{2}\phi'^{2} +
\frac{1}{2}m^{2}a^{2}\phi^{2} + \frac{1}{4}\lambda\phi^{4})}.
\label{eq:30}
\end{equation}
We clearly see that this is a good expression only if $a\ne0$. For the physical 
region $a > 0$ we can express the effective equation of state parameter as
\begin{equation}
w_{\rm{eff}} \equiv \frac{-\frac{1}{6}\phi'^{2} - \frac{1}{6}m^{2}a^{2}\phi^{2} 
+ \frac{1}{3}\rho_{r,0}}{-\frac{1}{2}\phi'^{2} + \frac{1}{2}m^{2}a^{2}\phi^{2} 
+ \rho_{r,0}}
\label{eq:31}
\end{equation}

\begin{figure}[t]
\begin{center}
\includegraphics[scale=0.67]{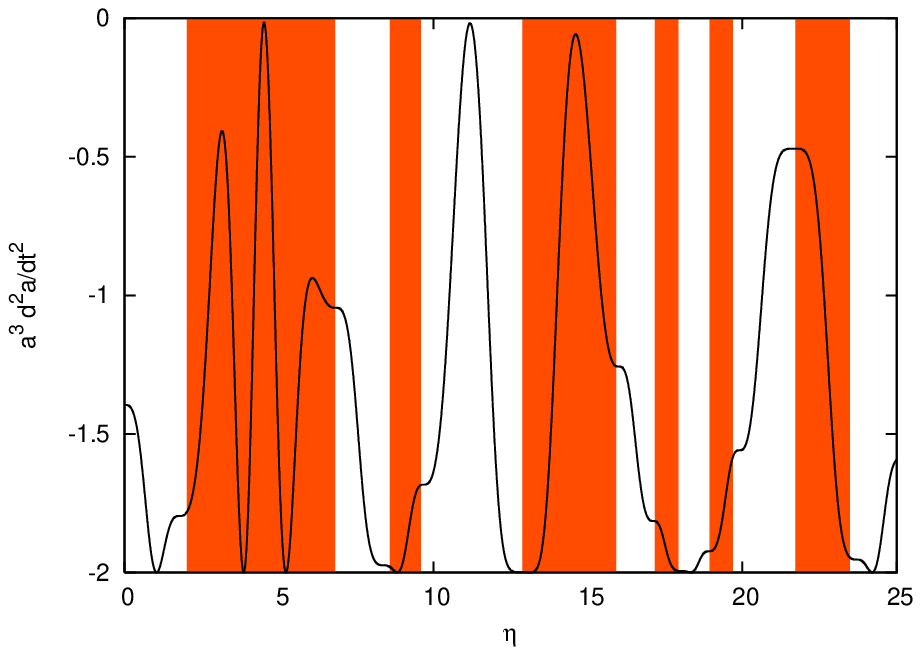}
\includegraphics[scale=0.67]{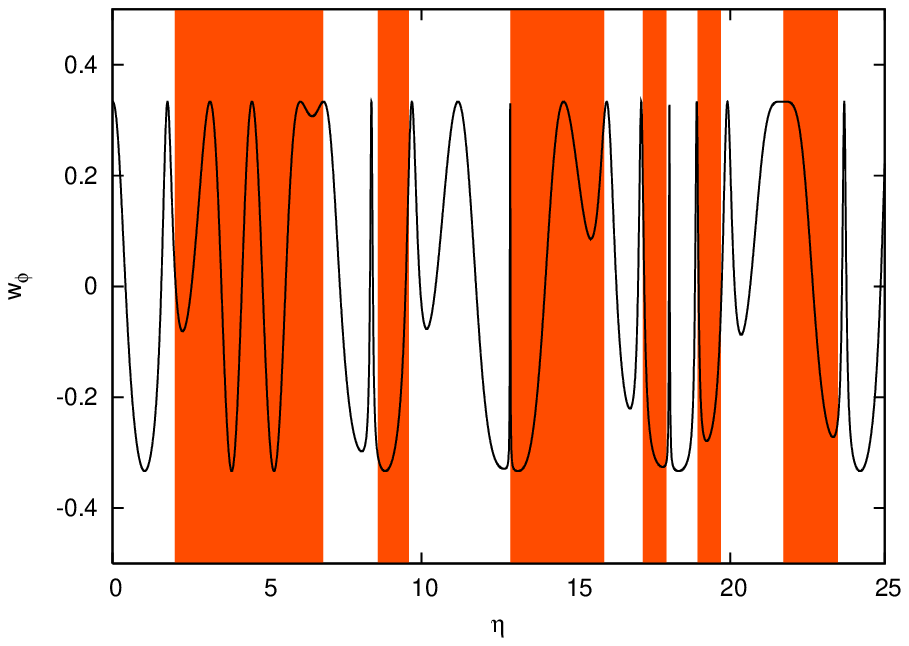}
\end{center}
\caption{The evolution of acceleration from Eq.~(\ref{eq:29}) with $\lambda=0$ 
(left panel) and equation of state parameter $w_\phi$ from Eq.~(\ref{eq:30}) 
(right panel) for model class I with respect to conformal time for short period 
of time for trajectory from left panel of Fig.~\ref{fig:1}. Shaded area denote
unphysical regions of scale factor $a \le 0$. In physical regions acceleration
is always negative $\ddot{a} \le 0$ model decelerates and equation of state
parameter never crosses $w_\phi=-1$ barrier.}
\label{fig:16}
\end{figure}

In Fig.~\ref{fig:16} we show the evolution of acceleration and evolution of 
coefficient of the equation of state $w_{\phi}$ for model class I (flat model
with $m^{2}<0$) with respect to the conformal 
time $\eta$ which is monotonous function of the cosmological time in the 
physical domain. Note that in the case of $m^2<0$ the universe is decelerating 
and $w_{\phi}$ stays in the interval $\langle -1/3 ; 1/3 \rangle$, with
$w_{\phi}=-1/3$ when $\phi'=0$ and $w_{\phi}=1/3$ when $\phi=0$.

\begin{figure}[t]
\begin{center}
\includegraphics[scale=0.67]{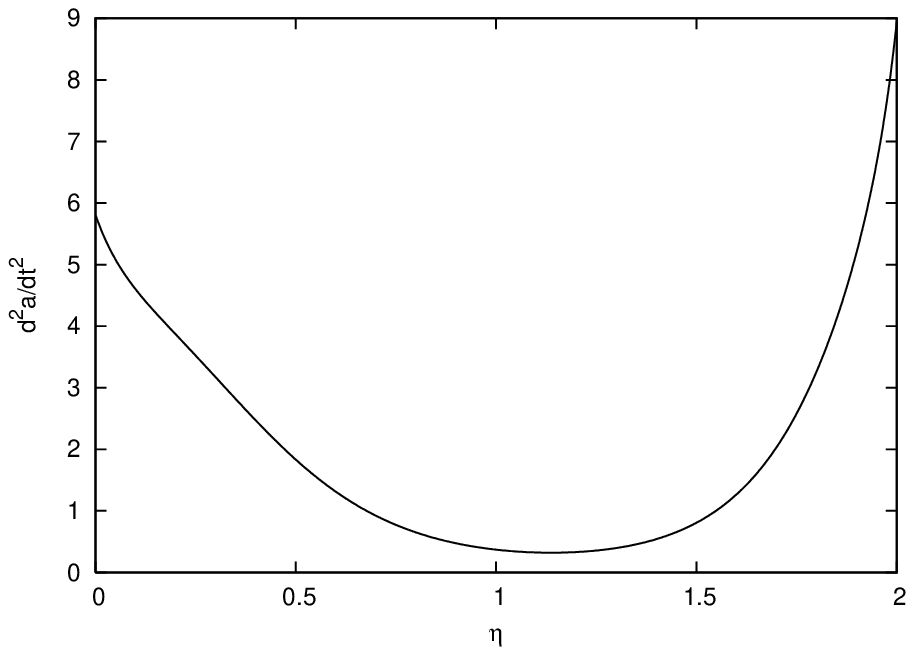}
\includegraphics[scale=0.67]{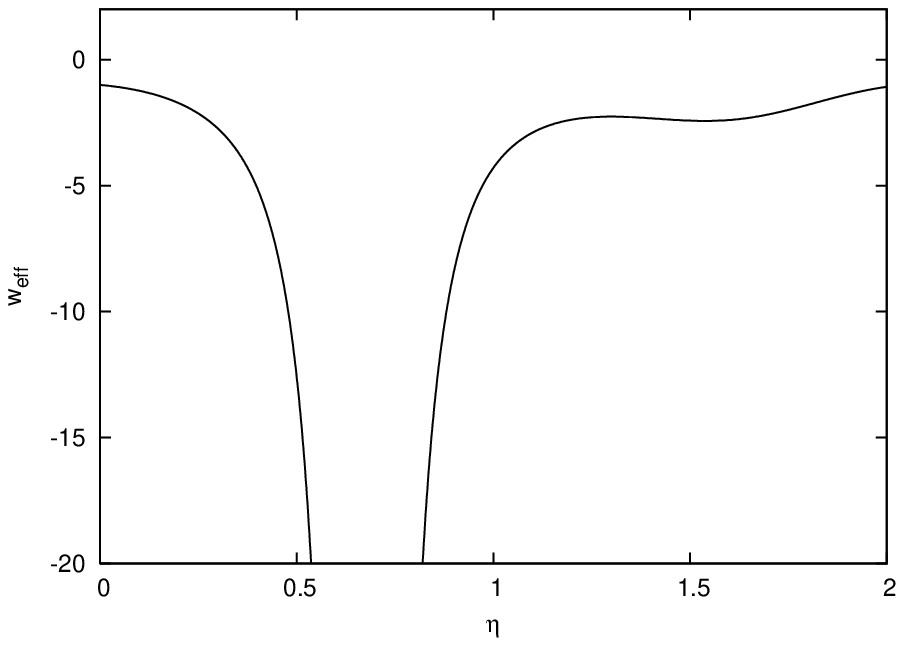}

\includegraphics[scale=0.67]{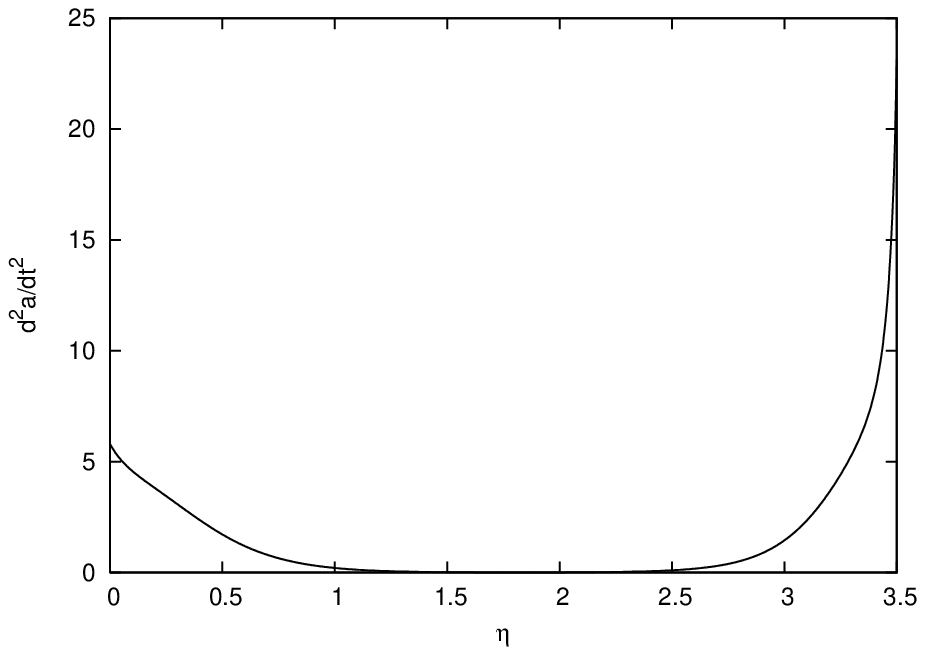}
\includegraphics[scale=0.67]{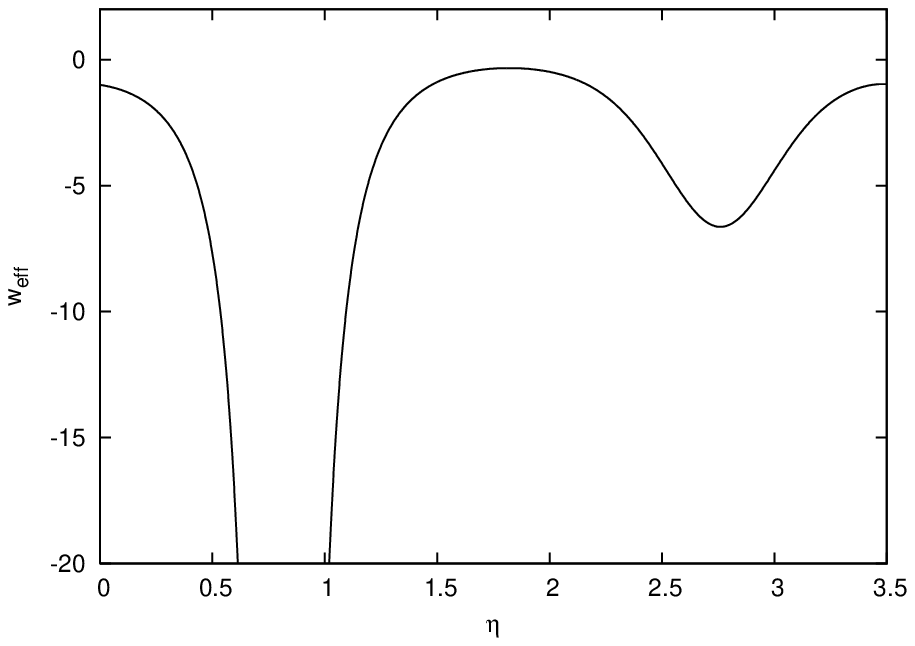}

\includegraphics[scale=0.67]{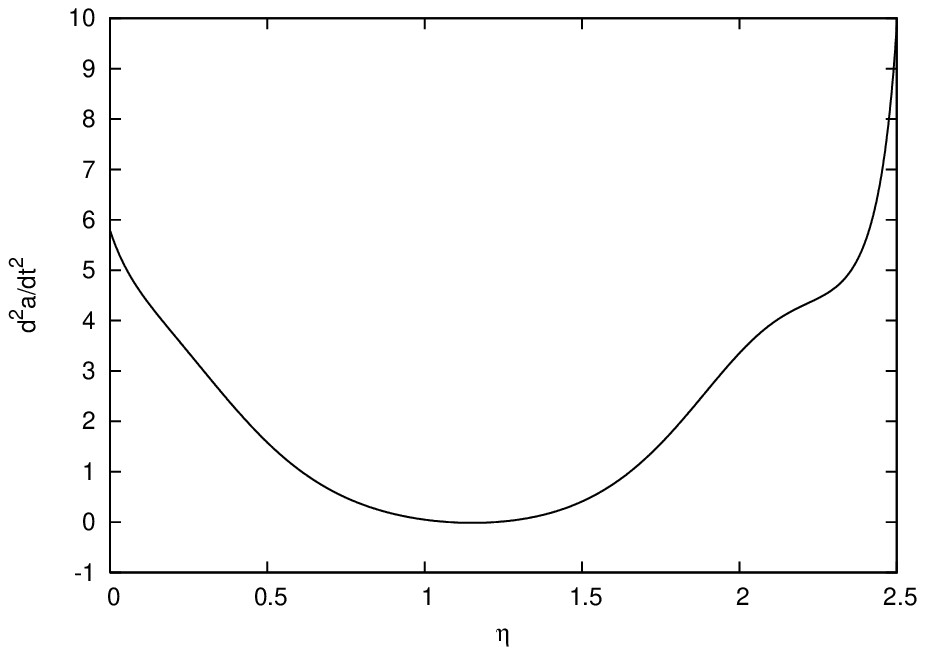}
\includegraphics[scale=0.67]{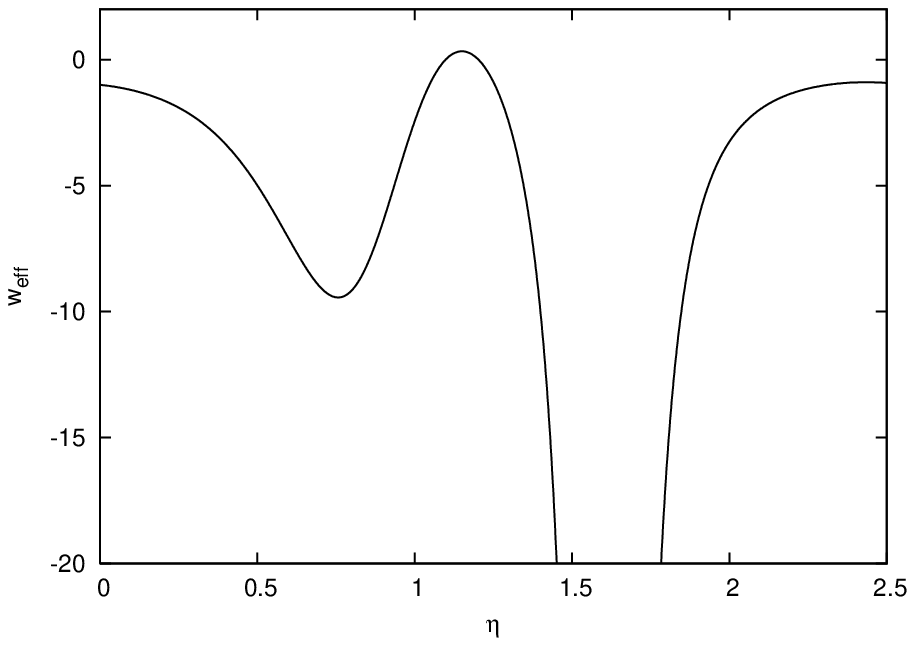}
\caption{The evolution of acceleration from Eq.~(\ref{eq:29}) (left panel) and
the equation of the state parameter $w_{\rm{eff}}$ from Eq.~(\ref{eq:31}) (right 
panel) for model class A without spontaneously symmetry breaking for a sample 
trajectory starting from the same initial conditions in the configuration space 
($a_{0}=4.75$, $\phi_{0}=-a_{0}+10$ and $a_{0}'=\phi_{0}'<0$ calculated from 
the Hamiltonian constrain) but for different energy levels:
$\mathcal{E}<0$ (top), $\mathcal{E}=0$ (middle) and $\mathcal{E}>0$ (bottom).
For every sample we have acceleration but the equation of the state parameter 
oscillates and aproaches $w_{\rm{eff}}=-1$.}
\label{fig:17}
\end{center}
\end{figure}

The analogous analysis made for the model class A without the spontaneously 
symmetry breaking (presented in Fig.~\ref{fig:17}) shows that the universe is 
always accelerating. In this case pressure $p_{\phi}$ is always negative. In 
the case of $\mathcal{E}>0$ we can have positive value of $w_{\phi}$ and this 
means that density can be negative. Moreover, during evolution there are 
crossings of the line $w_{\phi}=-1$ and then $w_{\rm{eff}}$ goes to minus one. 
The amplitude of these oscillations depends on $\rho_{\rm{r},0}$.

\section{Conclusions}

In cosmology the Mixmaster models are well known for their chaotic behavior 
\cite{Cornish:1995mf}. In these models chaos is caused by anisotropy of space. 
In this paper we considered the phantom cosmological models which are 
homogeneous and isotropic where the chaos is the consequence of the 
nonlinearity of the potential function for the scalar field. We found chaos is 
a generic feature of the phantom cosmology with the spontaneously symmetry 
breaking. It assumes the form of the multiple chaotic scattering process. In 
the case of the absence of symmetry breaking we did not find the sensitivity 
over initial conditions and the system behaves as the standard scattering 
process. 

There is no difference between the flat universes filled with minimal and 
conformally coupled scalar fields. As it was demonstrated by Faraoni et al. 
\cite{Gunzig:2000ce,Faraoni:2006sr} there is no chaos for the former class of 
models because the solution corresponding $k=0$ are restricted to move in 
some two-dimensional submanifold of the phase space $(H, \phi, \dot{\phi})$. 
There are no enough room for chaotic motion manifestation. The analogous result 
is valid for the other value of the coupling constant $\xi$. Note that we can 
observe chaotic behavior in some cases of the conformally coupled scalar field 
in the flat universes, when we consider the system on non-zero energy level. 
Physically it means that the universe is filled with radiation matter. 

The reason for which the flat universes with scalar field are non-chaotic 
lies in lack of invariant measure of chaotic behavior in the general 
relativity \cite{Szydlowski:1996uf,Motter:2003jm}. In general the 
standard chaos indicators like Lyapunov exponents or Kolmogorov-Sinai
entropy depends on time parameterization. In general relativity there 
is freedom of choice of the lapse function which defines reparameterization 
of time. The Hamiltonian is given modulo the lapse function. Therefore 
if we consider the Hamiltonian FRW system with scalar field on zero 
energy level it is possible that in the Hamiltonian constraint 
$a$ i $\dot{a}$ appear only in the combination $H=\dot{a}/a$. Hence 
the motion of the system takes place in the 3-dimensional phase space 
$h(H,\phi,\dot{\phi}) = 0$ on some 2-dimensional submanifold. For any 
value of coupling constant and general class of potentials there is 
no place for chaos (even the models with the cosmological constant). 

The new picture appears if the system is described on a non-zero energy 
level. Then the motion of the system is in the 3-dimensional submanifold 
of the 4-dimensional phase space. But in general case there is no possibility 
of choosing the lapse function in such a way that $a$ i $\dot{a}$ contribute 
in the Hamiltonian constraint by the Hubble function. 

Among the conformally coupled models there are in principal two types of 
behavior. First, chaotic scattering process takes place for case $m^2 < 0$ 
(the spontaneously symmetry breaking). In this case trajectories possess the 
property of topological transitivity which guarantees their recurrence 
in the phase space and the scattering process takes place around the origin. 
It is similar to the chaos appeared in the Yang-Mills theory. Exploring this 
analogy to this system we calculate a Gaussian curvature of the potential 
function which measure local instability of nearby trajectories. We proof 
that this curvature is negative. Because the configuration space is bounded 
in this case the mixing of trajectories is observed. 

Second, in the opposite case of $m^2 > 0$ we characterize dynamics in the system 
in terms of symbolic dynamics. No fractal structure in the space of initial 
conditions was found and conclude that the scattering process has not the 
chaotic character. 

We also found that the universe filled with conformally coupled scalar field 
is accelerating without the positive cosmological constant. In principle, 
there are two sources of this acceleration. The acceleration is driven by 
negative energy density and coefficient of equation of state $w_{\phi}$ is 
greater than $-1/3$. Alternatively, energy density is positive and the 
coefficient of equation of state can be smaller than $-1/3$. The 
latter there is crossing of $w_{\phi} = -1$. Of course, for the explanation 
of SNIa data apart from the acceleration of the Universe itself, the rate of 
acceleration is also crucial. 

The system with phantom scalar field coupled conformally to gravity in the 
FRW universe can be treated as scattering process with chaotic and nonchaotic 
character. The generic feature of this class of systems with the spontaneously 
symmetry breaking is the chaotic scattering of trajectories.

\begin{acknowledgments}
The paper was supported by the Marie Curie Actions Transfer of Knowledge 
project COCOS (contract MTKD-CT-2004-517186). We would like to thanks dr J. 
Grzywaczewski for his hospitality during the staying in Paris.
\end{acknowledgments}

\end{document}